\documentclass[aps,pra,amsmath,amssymb,floatfix,twocolumn,amsmath,superscriptaddress,twocolumn,nofootinbib,tighten,letterpaper]{revtex4-2}
\usepackage[colorlinks,linkcolor=blue,citecolor=blue,urlcolor=blue]{hyperref}
\usepackage{multirow}
\usepackage{subfigure}
\usepackage{color}
\usepackage{mathrsfs}
\usepackage{hyperref}
\usepackage[normalem]{ulem}
\usepackage{bm}

\usepackage{amssymb}   
\usepackage{amsmath}
\renewcommand\vec[1]{\ensuremath\boldsymbol{#1}} 

\usepackage{amsfonts, relsize, color}
\usepackage{graphics}
\usepackage{graphicx}
\usepackage{subfigure}
\usepackage{hyperref}
\usepackage{color}
\usepackage{comment}

\newcommand{\bra}[1]{\left< #1 \right|}
\newcommand{\ket}[1]{\left| #1 \right>}

\renewcommand\vec[1]{\ensuremath\boldsymbol{#1}} 

\begin{document}

\title{Projected branes as platforms for crystalline, superconducting, and higher-order topological phases}

\author{Archisman Panigrahi}\thanks{Contact author: archi137@mit.edu}
\affiliation{Department of Physics, Massachusetts Institute of Technology, Cambridge, Massachusetts 02139, USA}

\author{Bitan Roy}\thanks{Contact author: bitan.roy@lehigh.edu}
\affiliation{Department of Physics, Lehigh University, Bethlehem, Pennsylvania, 18015, USA}
\affiliation{Centre for Condensed Matter Theory, Department of Physics, Indian Institute of Science, Bengaluru 560012, India}

\date{\today}

\begin{abstract}
Projected branes are constituted by a \emph{small} subset of sites of a higher-dimensional crystal, otherwise placed on a hyperplane oriented at an irrational or a rational slope therein, for which the effective Hamiltonian is constructed by systematically integrating out the sites of the parent lattice that fall outside such branes [\href{https://doi.org/10.1038/s42005-022-01006-x}{Commun.\ Phys.\ {\bf 5}, 230 (2022)}]. Specifically, when such a brane is constructed from a square lattice, it gives rise to an aperiodic Fibonacci quasicrystal or its rational approximant (displaying emergent periodicity) in one dimension. In this work, starting from square lattice-based models for topological crystalline insulators, protected by the discrete fourfold rotational ($C_4$) symmetry, we show that the resulting one-dimensional projected topological branes encode all the salient signatures of such phases in terms of robust endpoint zero-energy modes, quantized local topological markers, and mid-gap modes bound to dislocation lattice defects, despite such linear branes lacking the $C_4$ symmetry of the original lattice. Furthermore, we show that such branes can also feature all the hallmarks of two-dimensional strong and weak topological superconductors through Majorana zero-energy bound states residing near their endpoints and at the core of dislocation lattice defects, besides possessing suitable quantized local topological markers. Finally, we showcase a successful realization of a square lattice-based second-order topological insulator with the characteristic corner-localized zero modes (protected by composites of nonspatial and crystalline symmetries) in its geometric descendant one-dimensional quasicrystalline or crystalline branes that generically feature a quantized localizer index, but support the endpoint zero-energy mode only at one of its endpoints when it passes through a corner of the parent crystal. Otherwise, topological endpoint modes at slightly higher energies (still separated from bulk states) appear at both ends of the brane, when they are slightly far from the square lattice corners. Possible engineered quantum and metamaterial-based platforms to experimentally harness our theoretically proposed topological crystalline and higher-order insulating as well as superconducting branes are discussed.   
\end{abstract}

\maketitle

\section{Introduction}

Classification of topological insulators and superconductors is primarily executed through the transformations of their representative effective single-particle Hamiltonian under \emph{three} nonspatial symmetry operations, namely (a) time-reversal symmetry, (b) (antiunitary) particle-hole symmetry, and (c) unitary particle-hole or sublattice or chiral symmetry~\cite{classification:1, classification:2, classification:3, classification:4, classification:5, classification:6, classification:7, classification:8, classification:9, classification:10, classification:11, classification:12, classification:13, classification:14, classification:15, classification:16, classification:17, classification:18, classification:19}. This procedure gives rise to tenfold Altland-Zirnbauer periodic table for topological phases of matter featuring numerous fascinating features, among which its Bott periodicity~\cite{classification:12} and dimensional reduction~\cite{classification:15} are the most prominent ones. Typically, Altland-Zirnbauer topological phases in $d$ dimension possess strong topological invariants which acquire quantized value over the corresponding $d$-dimensional Brillouin zone (BZ) in the reciprocal space. Nonetheless, besides strong topological phases, there also exist robust topological phases with net zero strong topological invariants, which on the other hand acquire nontrivial topological invariant, defined on lower-dimensional BZs. Such phases of matter are named weak topological phases~\cite{classification:10}.

Within the framework of the Altland-Zirnbauer classification scheme, once crystalline symmetries are invoked, the landscape of topological phases of matter immediately diversifies greatly, giving rise to topological crystalline insulators and superconductors. Crystalline topological phases display band inversion at multiple high-symmetry time-reversal invariant momentum (TRIM) points in the BZ, which are connected via discrete crystalline symmetries, such as rotations~\cite{FuTCI2011, Slager2012, Shiozaki2014}. In recent years, identification of crystalline topological phases (gapped and nodal) has given birth to topological quantum chemistry that nowadays is routinely employed to mine a variety of topological crystals in nature~\cite{bradlyn2017, PoVishwanath2017, slager2017, PoVishwanath2018, Fang2019, Vergniory2019, TangVishwanath2019}. 

\begin{figure}[t!]
	\includegraphics[width=1.00\linewidth]{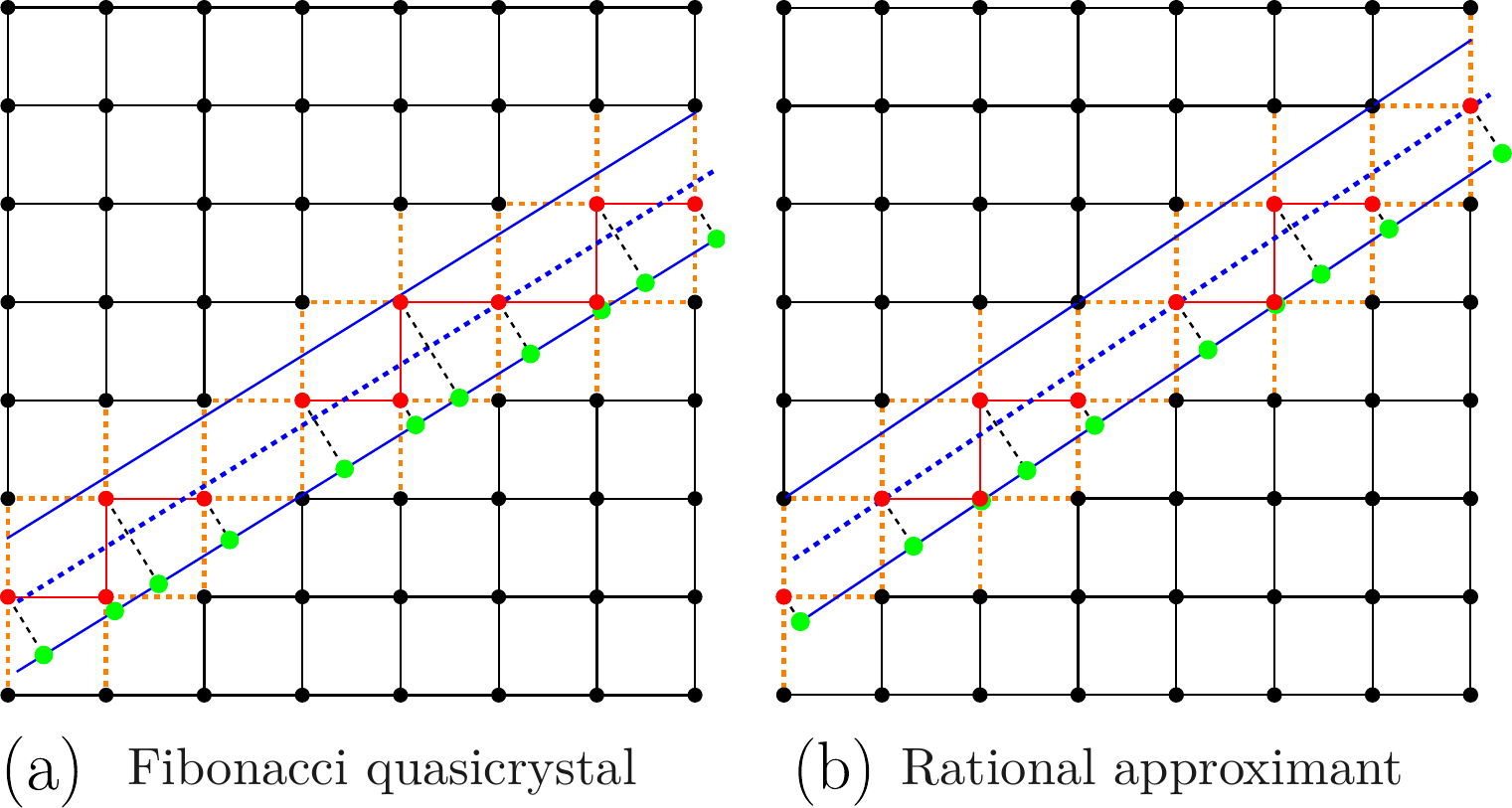}
	\caption{(a) Construction of a projected one-dimensional (1D) brane with an irrational slope of $m=(\sqrt{5}-1)/2$ (inverse of the golden ratio) from the parent square lattice. The black points represent sites in the region 2, and the solid black lines describe the hopping amplitudes in $H_{22}$. On the other hand, the red points represent the sites in the region 1 (lower-dimensional brane), bounded by the two blue lines $y = m x + c_{\rm up}$ and $y = m x + c_{\rm down}$ [see Eq.~\eqref{eq:lines}], and the red solid lines describe the hopping amplitudes in $H_{11}$. The orange dashed lines represent the hopping processes in the Hamiltonian of the parent square lattice system ($H_{\rm parent}$) among these two sets of sites and constitute $H_{12}$ and $H_{21}$. See Sec.~\ref{subsec:construction} and Eq.~\eqref{eq:Hparent} for details and notations. When the red colored sites are projected on the lower blue line, they constitute a 1D Fibonacci quasicrystal (green dots). The dotted blue line represents the central line within the 1D brane, with its equation given by $y = m x + (c_{\rm up} + c_{\rm down})/2$, on which we compute the local Chern marker, local weak invariant, and localizer index. (b) The rational approximant of the 1D brane with a rational slope of $m=2/3$. The colored lines and sites are analogous to those in panel (a). When the red colored sites are projected on the lower blue line (green dots), they constitute a rational approximant of the 1D Fibonacci quasicrystal displaying a periodic structure. For concreteness, here we only considered nearest-neighbor hopping processes to demonstrate the projection procedure. However, this procedure is applicable for any quadratic Hamiltonian $H_{\rm parent}$ on a parent square lattice, involving hopping amplitudes of arbitrary range.}
	\label{fig:lattice}
\end{figure}

About a decade ago, within the territory of topological phases of matter a new member emerged, which can be identified by the following observation. All the topological phases discussed so far when realized on $d$-dimensional crystals support robust gapless modes that reside on $(d-1)$-dimensional boundaries with $d_B=d-1$. Such boundary modes are also characterized by their codimension $d_c=d-d_B=1$, constituting the family of first-order topological phases. Now this concept can be generalized to envision topological phases hosting robust modes at the boundaries of a boundary which are characterized by $d_c=n$ (integer) with $n>1$. Corners with $d_c=d$ and hinges with $d_c=d-1$ are the prime examples of such lower-dimensional boundaries that can harbor topologically robust mid-gap modes. Namely, an $n$th order topological phase supports gapless modes on boundaries of codimension $d_c=n$, constituting the family of higher-order topological phases of matter that are typically protected by composites of nonspatial and crystalline symmetries~\cite{HOT:1, HOT:2, HOT:3, HOT:4, HOT:5, HOT:6, HOT:7, HOT:8, HOT:9, HOT:10, HOT:11, HOT:12, HOT:13, HOT:14, HOT:15, HOT:16, HOT:17}.

\subsection{Motivation, background, scope, and beyond}

Despite such diversity, nature only allows us to experimentally study topological materials up to three spatial dimensions, thereby limiting the exploration of higher dimensional topological phases. Therefore, in order to find the footprint of such topological phases within our three-dimensional world, we need an alternative approach to harness them. The recently proposed projected topological branes (PTBs) stand as promising candidates in this pursuit~\cite{PanigrahiPTB2022}. PTBs are constituted by a small subset of lattice sites within a higher-dimensional crystal that reside on a lower-dimensional hyperplane inclined at an irrational or a rational slope within the parent system, yielding either a quasicrystal or its rational approximant that shows emergent periodicity, which is, however, distinct from the one associated with the original lattice~\cite{QCGen:1, QCGen:2, QCGen:3}. Such a construction successfully captures all the signatures of square lattice-based strong topological insulators (featuring band inversion at the $\Gamma$ and $M$ points of the BZ with the latter one resulting from the translational symmetry of the square lattice) on one-dimensional PTBs (Fibonacci quasicrystals and their rational approximant)~\cite{PanigrahiPTB2022}. Furthermore, by stacking such one-dimensional PTBs in a translationally invariant fashion in the out of line direction, this construction successfully encodes the hallmarks of three-dimensional topological Weyl semimetals, such as the Fermi arcs, zeroth Landau level, and chiral anomaly in two dimensions. This construction has also been successfully extended to realize cubic lattice-based integer $Z$ topological insulators on a planar hexagonal brane~\cite{ TynerPTB2024} and identify square lattice-based strong, translationally active, and crystalline topological insulators on its geometric descendant Sierpi\'nski carpet fractal lattices, characterized by a noninteger Hausdorff dimension of $1.89$ (approximately)~\cite{salibmains2024}.

In this work, we extend the jurisdiction of PTBs to arrest higher-dimensional topological phases therein that are constrained by spatial symmetries and hence are more delicate, namely topological insulators protected by crystalline symmetries (topological crystalline insulators), and composites of crystalline and nonspatial symmetries (higher-order topological insulators) as well as topological superconductors (both strong and weak). Especially in the context of the former, the question of whether on a lower-dimensional brane we can find fingerprints of crystalline topological phases is of paramount fundamental importance as PTBs do not possess the crystalline symmetries of the original parent crystal. In this work, we present convincing numerical evidence, testifying to the successful realization of all these phases on one-dimensional PTBs starting from their parent square lattice-based prominent model Hamiltonian. Altogether, results presented here and in recent works~\cite{PanigrahiPTB2022, salibmains2024} establish the theoretical concept of PTBs as a promising avenue to harness predicted topological phases from four and higher dimensions within the landscape of designer quantum and metamaterials of dimensionality only up to three. These findings therefore should stimulate a new surge of experimental efforts geared toward controlled design of PTBs up to dimensionality of three to engineer even higher-dimensional topological phases therein. In this regard, quasicrystals are especially prominent as, for example, two-dimensional Penrose (three-dimensional icosahedral) quasicrystal is a holographic projection of five-dimensional (six-dimensional) hypercubic lattice~\cite{QCGen:2}.

 \subsection{PTBs: Construction and Hamiltonian}~\label{subsec:construction}

To summarize the central outcomes of this work in Sec.~\ref{subsec:summary}, we set the stage by promoting the general principle of constructing lower-dimensional projected branes from its parent higher-dimensional crystals. We denote the regions falling inside and outside the brane as region 1 and region 2, respectively. Let $H_{11}$ and $H_{22}$ be the Hamiltonian describing the hopping and onsite terms within these two regions, respectively, and $H_{12}$ and $H_{21}$ describe the inter-region hopping, with $H_{21} = H_{12}^\dagger$. Then the tight-binding Hamiltonian (or as such any quadratic Hamiltonian) of the whole parent system can be cast in the following block matrix form
\begin{equation}~\label{eq:Hparent}
    H_{\rm parent} = \left( \begin{array}{cc}
H_{11} & H_{12} \\
H_{21} & H_{22}
\end{array} \right).
\end{equation} 
The effective Hamiltonian for the PTB, denoted by $H_{\rm PTB}$, is then obtained by systematically integrating out the sites that fall outside such a brane in the path integral formalism, which is equivalent to computing the Schur complement of the sites within the brane~\cite{PanigrahiPTB2022}, leading to 
\begin{equation}~\label{eq:projected-Hamiltonian}
	H_{\rm PTB} = H_{11} - H_{12} H_{22}^{-1} H_{21}.
\end{equation}
In practice, we carry out this procedure in the following way. To construct one-dimensional (effectively) branes from a two-dimensional square lattice, we draw two straight lines governed by the equations
\begin{equation}~\label{eq:lines}
y = m \; x + c_{\text{up}} 
\:\: \text{and} \:\:
y = m \; x + c_{\text{down}},
\end{equation}
where $c_{\text{up}}$ and $c_{\text{down}}$ are two real numbers, and project the Hamiltonian for the parent system ($H_{\rm parent}$) to the sites contained between these two lines (see Fig.~\ref{fig:lattice}). For an irrational slope of $m$, the projected brane is a quasicrystal. Specifically, we focus on a Fibonacci quasicrystal, where the slope $m = (\sqrt{5}-1)/2$ is the inverse of the golden ratio. We also study the rational approximant of the associated quasicrystal, for which the slope is chosen to be $m = 2/3$ throughout this work. When the slope is rational, the brane displays a periodic structure.

By considering the effective tight-binding Hamiltonian ($H_{\rm parent}$) describing topological crystalline insulators (TCIs) for charged fermions and topological superconductors (TSCs) for neutral Bogoliubov-de Gennes (BdG) quasiparticles (encompassing both strong and weak topological phases) on the parent square lattice, in this work we search for their incarnations on the one-dimensional Fibonacci quasicrystal and its rational approximant crystalline branes from the spectral and topological properties of the corresponding $H_{\rm PTB}$. Furthermore, on such one-dimensional branes we also search for possible realizations of square lattice-based higher-order topological insulators (HOTIs), featuring zero-energy corner modes that are protected by combinations of nonspatial and crystalline symmetries. In this context, our main findings are summarized below.

\subsection{Summary of main results}~\label{subsec:summary}

Our current scientific voyage begins with a square lattice-based model Hamiltonian that supports fourfold rotational ($C_4$) symmetry protected time-reversal odd TCIs (namely, Chern insulators, hereafter named the ${\rm XY}$ phase). See the corresponding phase diagram in Fig.~\ref{fig:TCI-phase-diagram}(b). Then, following the general principle of constructing the effective Hamiltonian for one-dimensional PTBs, we numerically showcase the realization of such a crystalline ${\rm XY}$ phase therein. Specifically, we anchor this outcome from the topological zero energy modes at the end points of PTBs (Fig.~\ref{fig:TCI-LDOS-energy}), quantized local Chern markers of $\pm 2$ in the bulk of the central line of PTBs (Fig.~\ref{fig:TCI-local-marker}), and finally in terms of zero-energy modes bound to the dislocation core, which falls within the brane (Fig.~\ref{fig:TCI-dislocation-mode}), bearing the signature of the finite momentum band inversion associated with this phase in the parent square lattice.

Next we stop at the port of Majorana fermions and introduce the only momentum-independent local or on-site pairing allowed by the Pauli exclusion principle in a system that breaks the time-reversal symmetry in the normal state and composed of spinless or spin-polarized fermions. On a square lattice system, this paired state supports strong TSCs with Chern numbers ${\mathcal C}=\pm 2, \pm 1$ and a weak TSC with ${\mathcal C}=0$ but a quantized weak topological invariant, namely the Zak phase (Fig.~\ref{fig:SC-phase-diagram}). In this case, the projected one-dimensional superconducting brane supports zero-energy Majorana modes at its two ends in the parameter regime for the strong and weak TSCs (Fig.~\ref{fig:strong-topo-LDOS-energy}). While the strong TSCs, realized within PTBs, show quantized local Chern markers in the bulk of its central line that is equal to Chern number in parent system, the weak TSC lacks such a local strong topological marker (Fig.~\ref{fig:strong-topo-local-marker}). Instead the projected weak TSC displays a quantized local weak topological marker in the bulk of its central line (Fig.~\ref{fig:weak-topo-local-marker}). Finally, by noting that the strong TSCs with ${\mathcal C}=-2$ and $-1$ and the weak TSC feature inversion of the BdG bands at finite momentum on a square lattice BZ (namely, at the $M$ point), we search for dislocation bound Majorana zero modes within the projected branes when the core of such lattice defects falls within it. Intriguingly, we find that all such translationally active topological paired states foster defect bound Majorana zero-energy modes (Fig.~\ref{fig:topo-SC-dislocation-LDOS-energy}). We also note that when the local or on-site pairing term is introduced after projecting the Nambu-doubled normal state Hamiltonian from a square lattice to one-dimensional branes, its topological character remains unaffected (Fig.~\ref{fig:delta-mu-commuting-order}), in turn suggesting a robustness of the topological character in the paired states.

As the final topic of the current investigation, we focus on a member of the family of HOTIs on a square lattice that supports zero-energy topological modes that are localized at its four corners, named the second-order topological insulator (SOTI), for which quantized localizer index serves as the bona fide local topological invariant (Fig.~\ref{fig:HOTI-parent}). Following the general principle of constructing PTBs, when the square lattice-based model Hamiltonian for such a SOTI is projected onto one-dimensional branes, they continue to support zero-energy ($\sim 10^{-8} t$, where $t$ is the hopping amplitude) mode at one of its endpoints when it passes through one of the corners of the parent square lattice. Existence of such selective endpoint modes on one-dimensional PTBs corroborates with the quantized localizer index that can be found over a finite fraction of sites therein in the thermodynamic limit (Fig.~\ref{fig:HOTI-PTB}). By contrast, when none of the endpoints coincide with any corner of the square lattice, both the endpoints of one-dimensional branes can support topological modes, which are, however, placed at much higher energies ($\sim 10^{-2} t$). The variation of the localizer index with such an orientation of the brane within the parent lattice is similar to the scenario when one of its endpoints passes through a corner of the square lattice, as shown in Fig.~\ref{fig:HOTI-PTB-otherslope}. We find that irrespective of the slope of the brane and its relative position with respect to the corners of the parent square lattice, the localizer index assumes a trivial value in the entire brane when the parameter values represent a trivial insulator in the parent system, as shown in Fig.~\ref{fig:HOTI-PTB-trivial}.

In the only Appendix of this manuscript, we scrutinize the importance of the systematic site integration on the topological properties of PTBs. Specifically, we compute the local topological markers (strong and weak) only from $H_{11}$ and by completely ignoring the second term in $H_{\rm PTB}$, see Eq.~\eqref{eq:projected-Hamiltonian}. We find that neither the strong nor the weak local topological marker takes an integer-quantized value in this case (Fig.~\ref{app:fig:comparison-H11-HPTB}), thereby establishing the critical importance of the Schur complement or systematic site integration procedure to capture topological properties of higher-dimension crystals on their lower-dimensional projected branes.

\subsection{Organization}~\label{subsec:organization}

The remainder of the manuscript is organized as follows. Section~\ref{sec:projCTI} is devoted to the demonstration of two-dimensional square lattice-based TCIs on 1D PTBs in terms of robust end point near-zero energy modes, quantized local Chern markers, and mid-gap dislocation modes therein. Realizations of both two-dimensional strong and weak TSCs on 1D PTBs are staged in Sec.~\ref{sec:projTSC}. Realizations of planar SOTIs on 1D PTBs are shown in Sec.~\ref{sec:projHOTI}. In Sec.~\ref{sec:summary}, we summarize the central outcomes and present discussions on possible future directions and experimental realizations of the current study. Additional results, justifying the necessity of a systematic integration of the sites falling outside PTBs to construct their effective Hamiltonian are relegated to Appendix~\ref{app:comparison-H11-HPTB}.

\section{Projected crystalline topological insulators}~\label{sec:projCTI}

\begin{figure}[t!]
	\centering
	\includegraphics[width=1.00\linewidth]{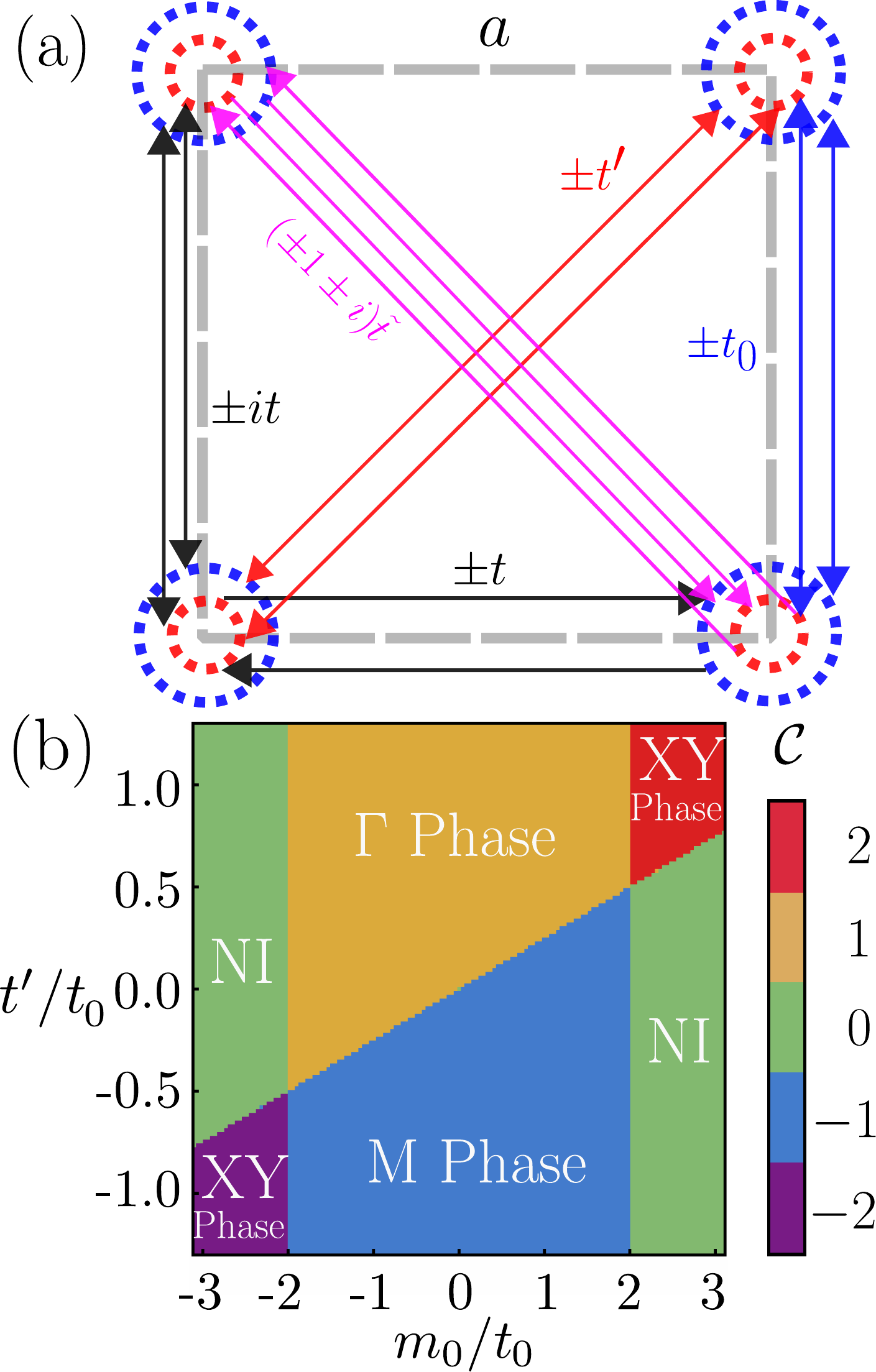}
	\caption{(a) A schematic of the hopping processes for the $\vec{d}$-vector in Eq.~\eqref{eq:dvectorTCI} between a few selected pairs of sites on a square lattice with lattice spacing $a$. On each site dashed blue and red circles correspond to orbitals with opposite parity eigenvalues. (b)Phase diagram of ${\mathcal H}_{\rm TCI}$, defined through Eqs.~\eqref{eq:TCI-Hamiltonian} and~\eqref{eq:dvectorTCI} at half-filling with $\tilde{t} = t =t_0= 1$ in terms of the integer Chern number (${\mathcal C}$) of the filled valence band. Topologically distinct insulating phases, named according to the band inversion momentum in the square lattice Brillouin zone, possess different integer values of ${\mathcal C}$. Inside the trivial or normal insulator (NI), ${\mathcal C}=0$. For details see Sec.~\ref{sec:projCTI}.}
	\label{fig:TCI-phase-diagram}
\end{figure}

We begin the discussion with a square lattice-based model Hamiltonian for crystalline symmetry protected Chern insulators and the corresponding phase diagram. While the existence of Chern insulators does not require any crystalline symmetry, there exist topological crystalline Chern insulators for which the band inversion simultaneously takes place at multiple TRIM points in the BZ that are connected by discrete crystalline symmetries of the underlying lattice, giving rise to robust gapless boundary or edge modes protected by a nontrivial bulk topological invariant~\cite{FuTCI2011, Slager2012, Shiozaki2014, bradlyn2017, PoVishwanath2017, slager2017, PoVishwanath2018, Fang2019, Vergniory2019, TangVishwanath2019}. A square lattice has discrete fourfold ($C_4$) crystalline rotational symmetry. Chern insulators protected by the $C_4$ symmetry can be described by the following minimal two-band Bloch Hamiltonian
\begin{equation}\label{eq:TCI-Hamiltonian} 
\begin{aligned}
    \mathcal{H}_{\rm TCI} = \sum_{\vec k} c^{\dagger}_{\vec k,\alpha} \: \left( H_{\rm TCI}(\vec k) \right)_{\alpha,\beta} \: c_{\vec k, \beta},
\end{aligned}
\end{equation}
where $H_{\rm TCI}(\vec k) =  \vec d(\vec k) \cdot \vec \tau$, $c^{\dagger}_{\vec k,\alpha}$ ($c_{\vec k,\alpha}$) denotes the fermionic creation (annihilation) operator with momentum ${\vec k}$ and on the orbital with parity eigenvalue $\alpha=+1$ or $-1$, and the set of two-dimensional Pauli matrices ${\vec \tau}=(\tau_x, \tau_y, \tau_z)$ operates on the orbital indices $\alpha$ and $\beta$. The components of the $\vec{d}({\vec k})$ vector are taken to be~\cite{Slager2012, dasroy2023}
\allowdisplaybreaks[4]
\begin{equation}~\label{eq:dvectorTCI}
    \begin{aligned}
        d_x (\vec{k}) & = t \sin (k_x a) + \tilde{t} \cos (k_x a) \sin (k_y a),  \\ 
        d_y (\vec{k}) & = t\sin (k_y a) + \tilde{t} \sin (k_x a) \cos (k_y a), \\ 
        \text{and} \:\: d_z (\vec{k}) & = m_0 - 2 t^\prime + t_0 [\cos (k_x a) + \cos (k_y a)] \\
                      & + 2 t^\prime \cos (k_x a) \cos (k_y a),
    \end{aligned}
\end{equation}
where $a$ is the lattice spacing of the square lattice. This model can readily be implemented on a square lattice by taking the Fourier transformation and we denote the corresponding real space Hamiltonian by $H^{\rm SL}_{\rm TCI}$, which is not shown here explicitly for brevity. Nonetheless, in Fig.~\ref{fig:TCI-phase-diagram}(a) we show a schematic of the hopping process with the above form of the $\vec{d}$-vector, where hopping amplitude between orbitals with same (opposite) parities along the principle axes is $t_0$ ($t$), and those along the body-diagonals is $t^\prime$ ($\tilde{t}$). The on-site staggered potential is given by $m_0-2 t^\prime$.

The phase diagram of this model can be obtained from the Chern number of the valence band of $H_{\rm TCI}(\vec k)$~\cite{TKNN1982}, for example, which we compute by discretization the BZ, following the method proposed by Fukui, Hatsugai, and Suzuki~\cite{Fukui2005}. The resulting phase diagram is displayed in Fig.~\ref{fig:TCI-phase-diagram}(b). To compute the Chern number of a given band, we divide the square-shaped BZ into an $N_k\times N_k$ square grid, where the point $(i,j)$ corresponds to momentum ${\vec k}_{i,j} = (k_i, k_j)$ with $k_i = 2\pi (i-1)/N_k-\pi$ and $i,j \in \{1,2,\cdots, N_k\}$. For the $n$th band, which has the Bloch wavefunction $\ket {n(\vec k_{i,j})}$ (a column vector), we define a $U(1)$ link variable as
\begin{equation}
U_{\mu}(\vec k_{i,j}) = \frac{\langle {n (\vec k_{i,j})}\ket{n (\vec k_{i,j} + {\vec \mu})}}{|\langle {n (\vec k_{i,j})} \ket{n (\vec k_{i,j} + {\vec \mu})}|}, 
\end{equation}
where $\vec \mu = 2\pi\hat {\vec e}_\mu/N$ with $\hat{\vec e}_\mu$ as the unit vector in the $x \; (y)$ direction for $\mu = 1 \; (2)$. The net (gauge invariant) Berry phase $\phi_{ij}$ acquired while traversing the counter-clockwise path around a unit square plaquette around the point $\vec k_{i,j}$ is given by
\begin{equation}
    e^{i\phi_{ij}} = U_1(\vec k_{i,j}) U_2(\vec k_{i,j} + \hat {\vec 1}) U_2(\vec k_{i,j} + \hat{\vec 2})^{-1} U_2(\vec k_{i,j})^{-1}.
\end{equation}
We then obtain the local lattice field strength tensor for the $n$th band $\tilde F_{12,n}(i,j) = \ln(\exp[i\phi_{ij}])$ with $-\pi < \tilde F_{12,n}(i,j)/i < \pi$. This quantity is the local Berry curvature multiplied by the area of the plaquette, yielding the Berry flux through the plaquette. Finally, the Chern number of the $n$th band (${\mathcal C}_n$) is obtained by summing the lattice field strength on each point of the discretized BZ lattice and is given by
\begin{equation}
    {\mathcal C}_n = \frac{1}{2\pi i} \sum_{i,j = 1}^{N_k} \tilde F_{12,n}(i,j).
\end{equation}
In the numerical computation we take $N_k=31$, which produces integer-quantized Chern numbers (within numerical accuracy). For the two-band Bloch Hamiltonian, defined through Eqs.~\eqref{eq:TCI-Hamiltonian} and~\eqref{eq:dvectorTCI}, we compute the Chern number of the only valence band and thus for notational simplicity we use ${\mathcal C}_n \equiv {\mathcal C}$ in Fig.~\ref{fig:TCI-phase-diagram}(b). When $\tilde{t} = t=1$, the system describes a topological insulator with the band inversion at the ${M}=(\pi,\pi)/a$ point of the BZ for $|m_0/t| < 2$ and $m_0 > t^\prime$. In this phase, known as the $M$ phase, the Chern number $\mathcal{C}=-1$ and the band inversion momentum ${\bf K}_{\rm inv}=(\pi,\pi)/a$. In contrast, the band inversion occurs at the $\Gamma = (0,0)$ point of the BZ when $|m_0/t| < 2$ and $t^\prime > m_0$ and the system is said to be in the $\Gamma$ phase with $\mathcal{C}=+1$ and ${\bf K}_{\rm inv}=(0,0)$. For $t^\prime > m_0 > 2t$, we find $\mathcal{C}=+2$ and when $t^\prime< m_0 < -2t$ we obtain $\mathcal{C} = -2$. In either of these two phases, the band inversion occurs simultaneously at the ${\rm X}=(\pi,0)/a$ and ${\rm Y}=(0,\pi)/a$ points of the BZ, which are connected by fourfold rotation, and the system is said to be in the ${\rm XY}$ or valley phase. In this work, we primarily focus on the ${\rm XY}$ phase. For other parameter values, this Hamiltonian describes a trivial insulator. 

\begin{figure*}[t]
	\centering
    \includegraphics[width=1.00\linewidth]{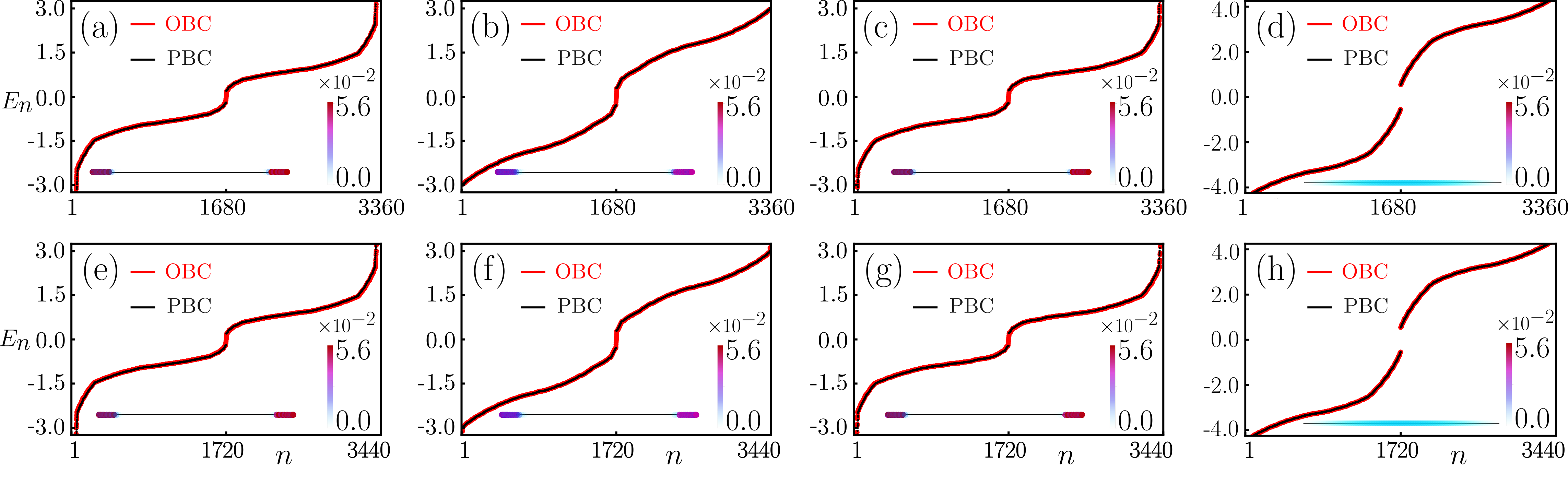}
	\caption{Energy spectrum of $H_{\rm PTB}$ [Eq.~\eqref{eq:projected-Hamiltonian}], obtained after projecting the square lattice-based tight-binding Hamiltonian $H^{\rm SL}_{\rm TCI}$ [with $H_{\rm parent} \equiv H^{\rm SL}_{\rm TCI}$ in Eq.~\eqref{eq:Hparent}], resulting from the Bloch Hamiltonian defined through Eqs.~\eqref{eq:TCI-Hamiltonian} and~\eqref{eq:dvectorTCI}, on an effective one-dimensional brane with an irrational slope of $m=(\sqrt{5}-1)/2$ within a $120 \times 120$ parent square lattice (yielding a Fibonacci quasicrystal) that contains only $11.66\%$ of its sites for (a) $t'=-1.5$ and $m_0=-3.0$ (${\rm XY}$ phase), (b) $t'=1.5$ and $m_0=0$ ($\Gamma$ phase), (c) $t'=1.5$ and $m_0=3.0$ (${\rm XY}$ phase), and (d) $t'=1.5$ and $m_0=-3.0$ (trivial phase). See Fig.~\ref{fig:TCI-phase-diagram}(b) for reference. Here, $E_n$ is the energy eigenvalue and $n$ is the eigenvalue index. Results are shown in red and black for open boundary condition (OBC) and periodic boundary condition (PBC) in each direction of the parent square lattice, respectively. In the inset of each corresponding subfigure, we show the probability density or local density of states for two states closest to zero energy in systems with OBCs for the same parameter values. Panels (e)--(h) are analogous to (a)--(d), respectively, but for a projected brane with a rational slope of $m = 2/3$ within a $120 \times 120$ parent square lattice (yielding a rational approximant of the Fibonacci quasicrystal), which contains only $11.94 \%$ of its sites. Throughout, we have set $c_{\rm down} = 4$ and $c_{\rm up} = 18$ [Eq.~\eqref{eq:lines}] and taken $t=\tilde{t}=t_0=1$. See Fig.~\ref{fig:lattice} for the construction of projected branes.}
	\label{fig:TCI-LDOS-energy}
\end{figure*}

A previous work has established the existence of the $\Gamma$ and $M$ phases on projected one-dimensional branes~\cite{PanigrahiPTB2022}. The existence of the $\Gamma$ phase does not rely on the underlying lattice structure and its signature can also be found on any noncrystalline system in two dimensions, such as fractal~\cite{noncryst:1, noncryst:2, noncryst:3} and amorphous~\cite{noncryst:4} lattices, and planar quasicrystals by directly implementing the models for two-dimensional topological insulators therein without any information regarding their parent systems~\cite{noncryst:5, noncryst:6, noncryst:7}. On the other hand, the $M$ phase results from the underlying translational symmetry of the square lattice, which can only be found on its geometric descendant Sierpi\'nski carpet fractal lattices~\cite{salibmains2024}, for example, besides on the one-dimensional PTBs~\cite{PanigrahiPTB2022}. Here, we are after a similar pursuit but for the valley or ${\rm XY}$ phases with ${\mathcal C} = \pm 2$ that are protected by the fourfold rotational symmetry of the square lattice. It would be interesting to investigate what happens to such valley or ${\rm XY}$ phases when the Hamiltonian is projected to one-dimensional branes, which do not have the $C_4$ symmetry of the parent square lattice. Interestingly, we find that the signatures of the ${\rm XY}$ phases, namely its bulk-boundary correspondence in terms of endpoint near zero-energy modes, quantized local Chern marker of $\pm 2$, as well as the presence of mid-gap zero-energy dislocation modes, survive even after the projection to a lower-dimensional brane, which we discuss in the following three sections, respectively.

\begin{figure}[t]
	\centering
	\includegraphics[width=1.00\linewidth]{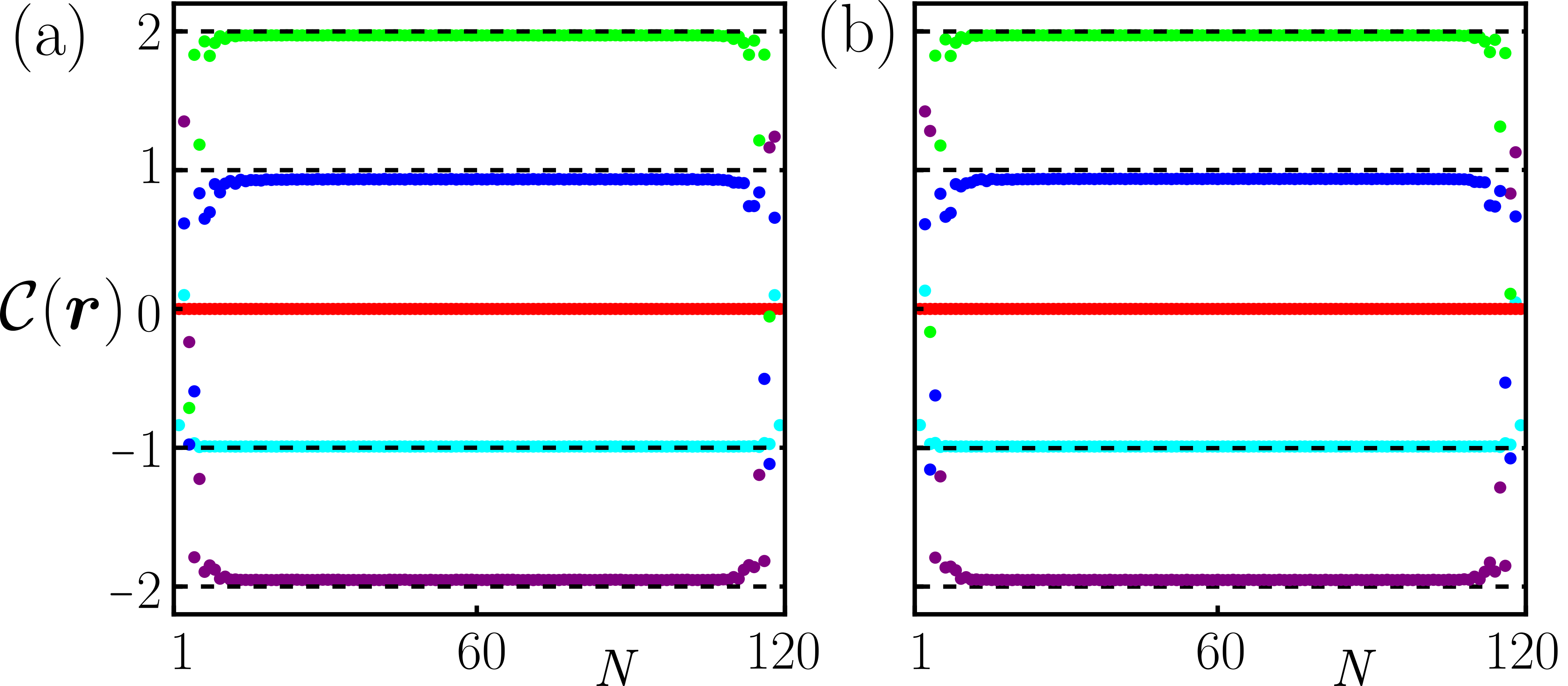}
	\caption{The local Chern marker $\mathcal{C}(\vec r)$ [Eq.~\eqref{eq:local-chern-definition}] of the projected topological  insulators (including the crystalline ones) along the central line (Fig.\ref{fig:lattice}) of the projected brane with (a) an irrational slope of $m=(\sqrt{5}-1)/2$ (yielding a Fibonacci quasicrystal) and (b) a rational slope of $m=2/3$ (resulting in a rational approximant of the quasicrystal) in a $120 \times 120$ parent square lattice, where we take $c_\text{down} = 4$ and $c_\text{up}=18$ [Eq.~\eqref{eq:lines}]. The green, blue, red, cyan, and purple curves are obtained for $(t^\prime,m_0) = (-1.5,-3)$, $(-1.5,0)$, $(1.5,-3)$, $(0.5,0)$, and $(1.5,3)$, respectively. For these parameter values the parent square lattice model supports the ${\rm XY}$ phase, $\Gamma$ phase, trivial phase, $M$ phase, and ${\rm XY}$ phase, respectively. For the corresponding quantized Chern number see Fig.~\ref{fig:TCI-phase-diagram}(b). For the irrational (rational) slope, the projected brane contains only $11.66\%$ ($11.94\%$) sites of the parent square lattice. Throughout, we set $t=\tilde{t}=t_0=1$. Here, $N$ is the site number on the projected topological brane along its central line [Fig.~\ref{fig:lattice}].
    }~\label{fig:TCI-local-marker}
\end{figure}

\subsection{TCIs: from square lattice to 1D branes}

To demonstrate that the signatures of the ${\rm XY}$ phase survive after projecting the square lattice-based Hamiltonian to  effective one-dimensional branes, we first diagonalize the corresponding $H_{\rm PTB}$ from Eq.~\eqref{eq:projected-Hamiltonian} and search for near zero-energy modes within the topological parameter regime of the parent Hamiltonian $H_{\rm parent} \equiv H^{\rm SL}_{\rm TCI}$ in Eq.~\eqref{eq:Hparent} and subsequently scrutinize the profile of their local density of states (LDOS). We find that close to the zero energy eigenstates of the projected Hamiltonian ($H_{\rm PTB}$) are quantitatively similar to that of the parent Hamiltonian on a square lattice ($H^{\rm SL}_{\rm TCI}$) in the following way. Specifically, we find that in the parameter regimes where the parent square lattice hosts topologically nontrivial phases ($\Gamma$, $M$, and valley or ${\rm XY}$ phases) that support one-dimensional zero-energy edge modes, the projected Hamiltonian accommodates near zero-energy gapless modes that are localized near the end points of the one-dimensional PTBs under open boundary conditions (OBCs) in both directions. These outcomes on the Fibonacci quasicrystal are shown in Figs.~\ref{fig:TCI-LDOS-energy}(a)--\ref{fig:TCI-LDOS-energy}(c), and on its rational approximant are shown in Figs.~\ref{fig:TCI-LDOS-energy}(e)--\ref{fig:TCI-LDOS-energy}(g), with the corresponding LDOS shown in the respective inset. Therefore the endpoint modes of PTBs are topological holographic images of the gapless edge modes localized near the boundaries of the parent square lattice. Under periodic boundary conditions (PBCs), the spectrum of $H_{\rm PTB}$ remains gapped, as expected. For the trivial insulator, the spectrum remains gapped under both PBC and OBC, and the LDOS for two closest to zero-energy modes is spread throughout the bulk of the PTBs, as shown in Fig.~\ref{fig:TCI-LDOS-energy}(d) and~\ref{fig:TCI-LDOS-energy}(i) for Fibonacci quasicrystal and its rational approximant, respectively. Next we compute a nontrivial topological invariant on such PTBs that protect such end-point zero-energy modes.

\subsection{Local Chern marker on 1D projected brane}~\label{subsec:LCMTCI}

The topological invariant of crystalline topological insulator described by $H_{\rm TCI}(\vec{k})$ is the Chern number, which is defined as the integral of the Berry curvature over the BZ~\cite{TKNN1982, Fukui2005}, which we have discussed in the previous subsection to arrive at the phase diagram shown in Fig.~\ref{fig:TCI-phase-diagram}(b). The real-space analog of the Chern number is the Bott index \cite{Hastings2010a, Hastings2010b, Toniolo2022}. To calculate the Bott index, one needs to utilize the fact that the system has a translational invariance in two dimensions. As the effective one-dimensional brane lacks two-dimensional translation invariance, we cannot directly calculate the (global) Chern number or the Bott index of its bands. Instead, we have to rely on local topological markers to calculate the topological invariant of the phases described by the projected Hamiltonian $H_{\rm PTB}$. In this case, the appropriate local topological marker is the local Chern marker~\cite{Bianco2011, Chen2023UniversalTopoMarker, MannaDasRoy2024, salibroy:LCM}, which is contained within the operator
\begin{equation}\label{eq:local-chern-marker}
	\mathcal{L}_{\text{Chern}} = -4\pi \; {\rm Tr}\left[\rm{Im}\{ \hat{P} \hat{X} \hat{Q} \hat{Y} \hat{P} \}\right].
\end{equation}
Here, $\hat{P}$ is the projector onto the valence band, its complimentary operator $\hat{Q} = 1 - \hat{P}$ projects onto the conduction band, $\hat X$ and $\hat Y$ are the position operators in the $x$ and $y$ directions, respectively, and `${\rm Tr}$' denotes the trace over the orbital degrees of freedom. The local Chern marker on a site at position ${\vec r}$ is the expectation value of the matrix operator $\mathcal{L}_{\text{Chern}}$ at ${\vec r}$, given by
\begin{equation}~\label{eq:local-chern-definition}
    \mathcal C(\vec r) = \bra {\vec r}\mathcal{L}_{\text{Chern}} \ket {\vec r}.
\end{equation}
Since we use the real space representation of the Hamiltonian, $\mathcal C(\vec r)$ corresponds to the diagonal entries of $\mathcal{L}_{\text{Chern}}$. 

\begin{figure}[t!]
	\centering
	\includegraphics[width=1.00\linewidth]{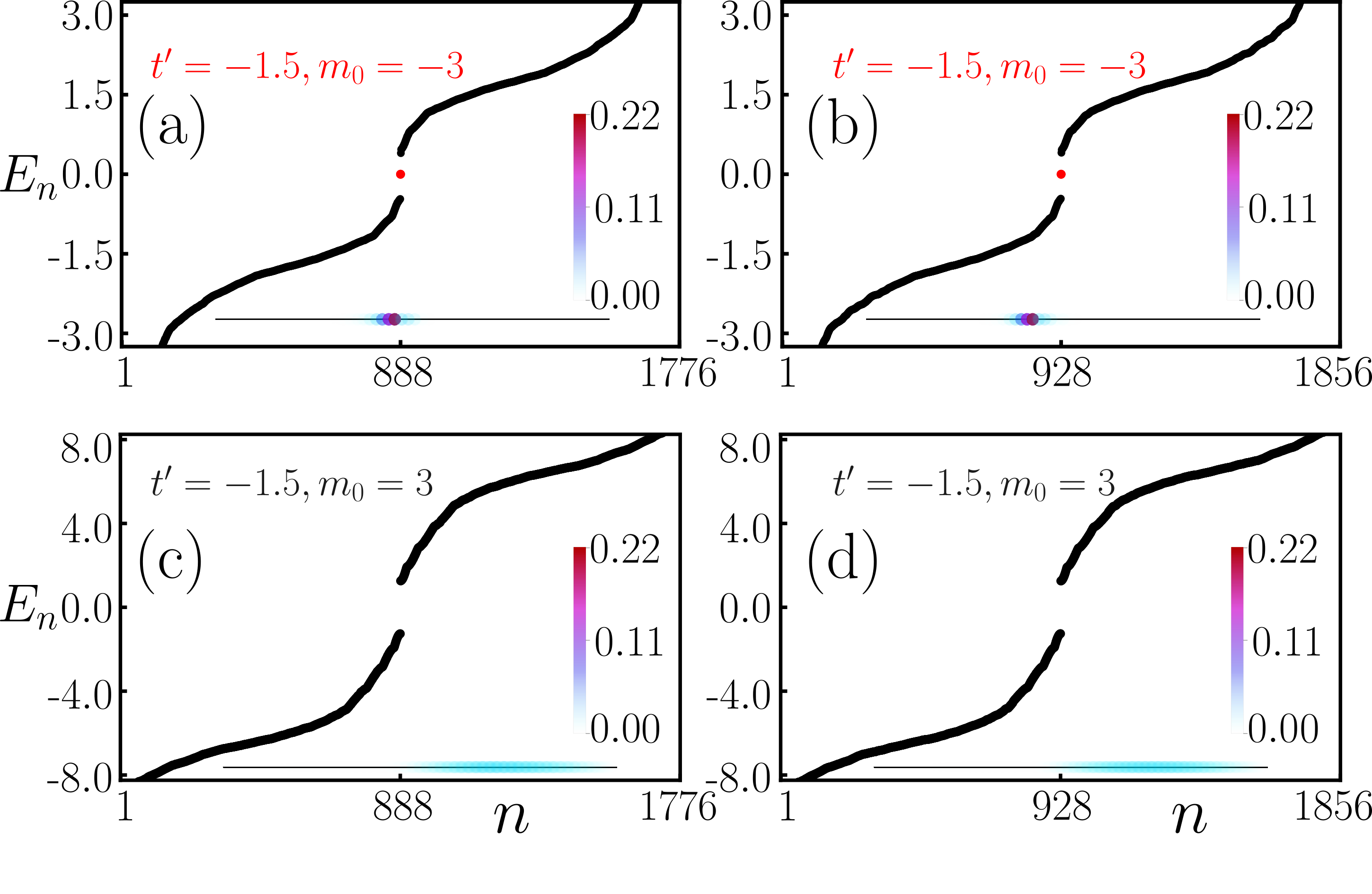}
	\caption{Dislocation modes of projected crystalline topological insulators. (a) Energy spectra of $H_{\rm PTB}$ in a projected Fibonacci quasicrystalline brane with a dislocation core falling within it for $t^\prime = -1.5$ and $m_0 = -3$ in $H^{\rm SL}_{\rm TCI}$, thereby yielding the ${\rm XY}$ phase on the parent square lattice, see Fig.~\ref{fig:TCI-phase-diagram}(b). Despite a projection to the brane, the system continues to host localized dislocation modes very close to the zero-energy (shown in red) that are separated from the other bulk states (shown in black), as the parent system is in the translationally active ${\rm XY}$ phase. The inset shows the probability density of the dislocation modes, confirming their sharp localization around the defect core. Here, $E_n$ is the energy eigenvalue and $n$ is the eigenvalue index. (b) Analogous to (a), but for a projected brane that is a rational approximant of the Fibonacci quasicrystal. (c) The energy spectra of the trivial phase for $t' = -1.5$ and $m_0 = 3$ in the same setup. Here, the states closest to zero energy are spread throughout the bulk of the quasicrystalline brane as confirmed from their probability density shown in the inset. (d) Analogous to (c), but for the rational approximant brane of the Fibonacci quasicrystal. In every plot, we utilize a $100 \times 100$ parent square lattice with a single edge dislocation with the Burgers vector ${\bf b} =a {\hat x}$ and periodic boundary condition in the $x$ direction, and project the corresponding $H^{\rm SL}_{\rm TCI}$ to one-dimensional branes with $c_\text{down} = -12$ and $c_\text{up}=-2$ [Eq.~\eqref{eq:lines}]. Then the projected Fibonacci brane with $m = (\sqrt{5}-1)/2$ contains $8.83\%$ of the sites and its rational approximant with $m=2/3$ contains $9.23\%$ of the sites of the parent square lattice. Throughout, we set $\tilde{t} = t= t_0 =1$.}
	\label{fig:TCI-dislocation-mode}
\end{figure}

We obtain the local Chern marker on PTBs by utilizing the projection matrices $\hat P$ and $\hat Q$ of the bands of the projected Hamiltonian $H_{\rm PTB}$. For the position operators in this case, we use the coordinates of the sites belonging to the branes in the original coordinate system of the parent square lattice. For every $x$ coordinate, we identify the site closest to the central line (blue dashed lines in Fig.~\ref{fig:lattice}), determined by the equation [see also Eq.~\eqref{eq:lines}] 
\begin{equation}~\label{eq:centralline}
y = mx + \frac{c_{\rm up} + c_{\rm down}}{2}
\end{equation}
and compute the local Chern marker for $H_{\rm PTB}$ on this set of sites. Even though the effective one-dimensional brane does not have a fourfold rotational symmetry of the parent square lattice, the parameter values over which the parent system supports the valley or ${\rm XY}$ phases with Chern numbers of ${\mathcal C}=\pm 2$, the PTBs continue to display a quantized local Chern number $\pm 2$ on the central line of the quasicrystalline brane and its rational approximant, as shown in Figs.~\ref{fig:TCI-local-marker}(a) and~\ref{fig:TCI-local-marker}(b), respectively. Therefore, the local Chern marker in the PTB is consistent with the Chern number or the Bott index for the parent square lattice model. Hence, the near zero-energy end point modes (see Fig.~\ref{fig:TCI-LDOS-energy}) on the PTBs are protected by quantized local Chern marker, assuring the topological incarnation of the valley or ${\rm XY}$ phase therein. It should, however, be noted that near the ends of the PTBs, the local Chern marker deviates from its quantized values, which is insensitive to the boundary conditions as the position operators ${\hat X}$ and ${\hat Y}$ are not periodic. A similar boundary effect exists for the local Chern marker on the parent square lattice~\cite{salibroy:LCM}.

\subsection{Dislocation modes of projected TCIs}~\label{subsec:DislocationTCI}

\begin{figure*}[t]
	\centering
	\includegraphics[width=1.00\linewidth]{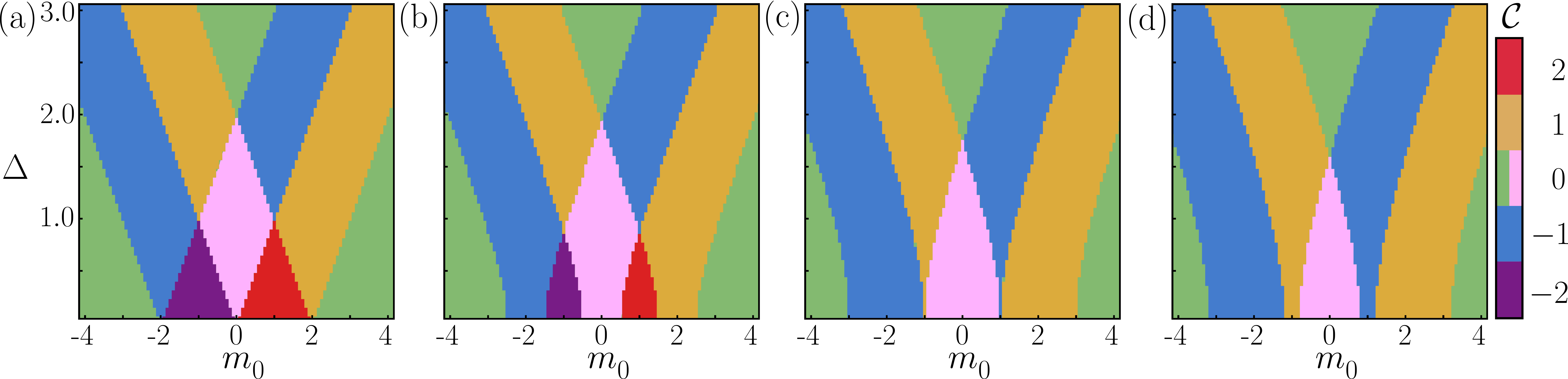}
	\caption{Phase diagrams of the Bogoliubov-de Gennes Hamiltonian from Eq.~\eqref{eq:topo-SC-Hamtiltonian} with (a) $\mu=0.0$, (b) $ 0.5$, (c) $ 1.0$, and (d) $1.2$ in terms of the total Chern number of the two valence bands of neutral Majorana fermions (${\mathcal C}$) on the $(m_0, \Delta)$ plane, where $\Delta$ is the pairing amplitude and $\mu$ is the chemical potential. For $\mu \geq 1.0$ the phases with $\mathcal{C}=\pm2$ are completely suppressed. The pink shaded region marks the weak topological superconductor, which features a nonzero weak topological invariant despite having a net zero total Chern number. In this phase, while the two bands below the Fermi level are individually topological, their Chern numbers cancel each other. Throughout, we set $t=t_0=1$. For details see Sec.~\ref{sec:projTSC}.
    }
	\label{fig:SC-phase-diagram}
\end{figure*}

The topological nature of an insulator, specifically the associated band inversion momentum ${\bf K}_{\rm inv}$, can be identified by studying the mid-gap dislocation modes at zero energy that emerge when the underlying two-dimensional lattice hosts edge dislocation, which locally breaks the translational symmetry. For example, if the band inversion occurs at zero momentum, as is the case in the $\Gamma$ phase where ${\bf K}_{\rm inv}=(0,0)$, the system does not host any zero-energy modes localized at the dislocation core. But such modes exist at zero-energy when the band inversion occurs at a finite momentum that together with the Burgers vector ${\bf b}$, characterizing the dislocation lattice defect, satisfy the quantization condition ${\bf K}_{\rm inv} \cdot {\bf b}=\pi$ (modulo $2 \pi$)~\cite{defect:1, defect:2, defect:3, defect:4, defect:4a, defect:5, defect:6, defect:6a, defect:7, defect:8, defect:9, defect:10, defect:11, defect:12,defect:13,defect:14, defect:15, defect:16, defect:17, defect:18, defect:19, defect:20, defect:21, defect:22}, placing the system in a translationally active phase, as is the case for $M$ and ${\rm XY}$ phases.

To establish such a robust ${\bf K}\cdot {\bf b}$ rule, consider an edge dislocation with Burgers vector $a \; \hat{j}$, where $\hat j$ is a unit vector along the $j$th direction. An electron with momentum $\vec p_0$ that encircles an edge dislocation accumulates a net phase factor of $\exp\left[i a \hat j \cdot \vec p_0 \right]$ with $\vec p_0 ={\bf K}_{\rm inv}$ in a topological insulator. Consequently, for a topological insulator in the $\Gamma$ phase, the acquired phase is zero and the electrons do not respond to dislocation lattice defects. By contrast, in the other topological phases involving a finite momentum band inversion, namely the $M$ phase with ${\bf K}_{\rm inv}=(\pi,\pi)/a$ and ${\rm XY}$ phases with ${\bf K}_{\rm inv}=(\pi,0)/a$ and $(0,\pi)/a$ (simultaneously), the electrons gain a topologically nontrivial $\pi$ phase while encircling the edge dislocation with ${\bf b}=a {\hat x}$ or $a {\hat y}$. Massive Dirac fermions nontrivially respond to such $\pi$-flux, threading through the dislocation core and harbor mid-gap zero-energy modes that are highly localized around the defect cores~\cite{defect:3}.

Here we demonstrate that the original dislocation modes of the translationally active ${\rm XY}$ phase on the parent square lattice are robust and survive both in the Fibonacci quasicrystal and its rational approximant. Specifically, in the construction of the PTBs we choose $c_{\rm up}$ and $c_{\rm down}$ such that the dislocation center of the parent square lattice falls within the projected brane~\cite{PanigrahiPTB2022}. We then find that the projected brane hosts zero-energy dislocation modes as long as the parent system is in a translationally active ${\rm XY}$ phase. Specifically, in Fig.~\ref{fig:TCI-dislocation-mode} we display the eigenvalue spectrum of $H_{\rm PTB}$, obtained by projecting the corresponding square lattice Hamiltonian supporting a single edge dislocation with ${\bf b}=a {\hat x}$ and periodic boundary condition only in the $x$ direction such that the parent system is in the translationally active valley ${\rm XY}$ phase. Indeed, we find modes that are sufficiently close to the zero-energy and the corresponding LDOS confirms their high localization around the defect core within the PTB, as shown in the insets of Fig.~\ref{fig:TCI-dislocation-mode}. In turn, this observation guarantees the translationally active nature of the crystalline symmetry protected ${\rm XY}$ phase on the PTBs, featuring either a quasicrystal or its rational approximant. However, such modes are absent when the system is a trivial insulator (also shown in Fig.~\ref{fig:TCI-dislocation-mode}) and $\Gamma$ phase. Existence of such topological dislocation modes within the PTBs for the $M$ phase of the parent square lattice has already been shown in Ref.~\cite{PanigrahiPTB2022}, which we thus do not display here explicitly.

\section{Projected Topological Superconductor}~\label{sec:projTSC}

In this section, we show that projected one-dimensional branes can also harbor both two-dimensional strong and weak topological thermal insulators or superconductors, realizable on the parent square lattice. To capture these phases, we consider the following effective single-particle BdG Hamiltonian for neutral Majorana fermions of the following form on a square lattice-based system, in which the pairing occurs between spinless fermions or spin-polarized electrons or holes living on the same site but on orbitals with opposite parities
\begin{equation}
    \mathcal{H}_{\rm BdG} = \frac{1}{2} \sum_{\vec k} \Psi_{\vec k}^\dagger \; H_{\rm BdG}(\vec k) \; \Psi_{\vec k},
\end{equation}
where the Nambu-doubled spinor basis reads as 
\begin{equation}
\Psi^\top_{\vec k} = (c_{\vec k, +}, c_{\vec k, -}, c^\dagger_{-\vec k, -}, c^\dagger_{-\vec k, +}),    
\end{equation}
and the subscripts $\pm$ denote the parity eigenvalues of the orbitals. The factor of $1/2$ accounts for the Nambu doubling. As an instructive toy model for TSCs, displaying both strong and weak topological behaviors, we consider the following $4\times4$ effective single-particle Bloch-BdG Hamiltonian in the announced Nambu-doubled basis~\cite{Das2023} 
\begin{equation}~\label{eq:topo-SC-Hamtiltonian}
	\begin{aligned}
	H_{\rm BdG}(\vec k) &= t \sin (k_x a) \Gamma_{01} + t \sin (k_y a) \Gamma_{02} + \Delta \Gamma_{13}  \\ 
                        &+ [m_0 - t_0 (\cos (k_x a) + \cos (k_ya) )] \Gamma_{03}  - \mu \Gamma_{30}, 
	\end{aligned}
\end{equation} 
where $\Delta$ is the pairing amplitude and $\mu$ is the chemical potential, measured from the zero-energy (band center). Four-dimensional Hermitian Dirac matrices are defined as $\Gamma_{ij} = \eta_i \otimes \tau_j$, where two sets of Pauli matrices $\{ \eta_\rho \}$ and $\{ \tau_\rho \}$ with $\rho=0,1,2,3$ operate on the particle-hole or Nambu and orbital degrees of freedom, respectively, and $\otimes$ is the Kronecker product. In the absence of any chemical potential and pairing, the resulting two-dimensional Bloch Hamiltonian for the normal state (captured solely in terms of the Pauli matrices $\{ \tau_\rho \}$) describes the Qi-Wu-Zhang~\cite{Qi-Wu-Zhang2006} or spinless version of the Bernevig-Hughes-Zhang~\cite{classification:7} model for quantum anomalous and time-reversal symmetry-breaking normal insulators in the parameter regimes $|m_0/t_0|<2$ and $|m_0/t_0|>2$, respectively. The corresponding Hamiltonian takes the generic form shown in Eq.~\eqref{eq:TCI-Hamiltonian} with the components of the $\vec{d}(\vec{k})$-vector obtained from Eq.~\eqref{eq:dvectorTCI} after setting $\tilde{t}=t^\prime=0$ therein. Such a system can permit only one local pairing corresponding to the number of purely imaginary Hermitian two-dimensional matrices to comply with the Pauli exclusion principle. Once again we implement $H_{\rm BdG}(\vec k)$ on a square lattice after its Fourier transformations to the real space, which we denote by $H^{\rm SL}_{\rm BdG}$. For brevity, here we do not show $H^{\rm SL}_{\rm BdG}$ explicitly.  

\begin{figure*}[t]
	\centering
    \includegraphics[width=0.95\linewidth]{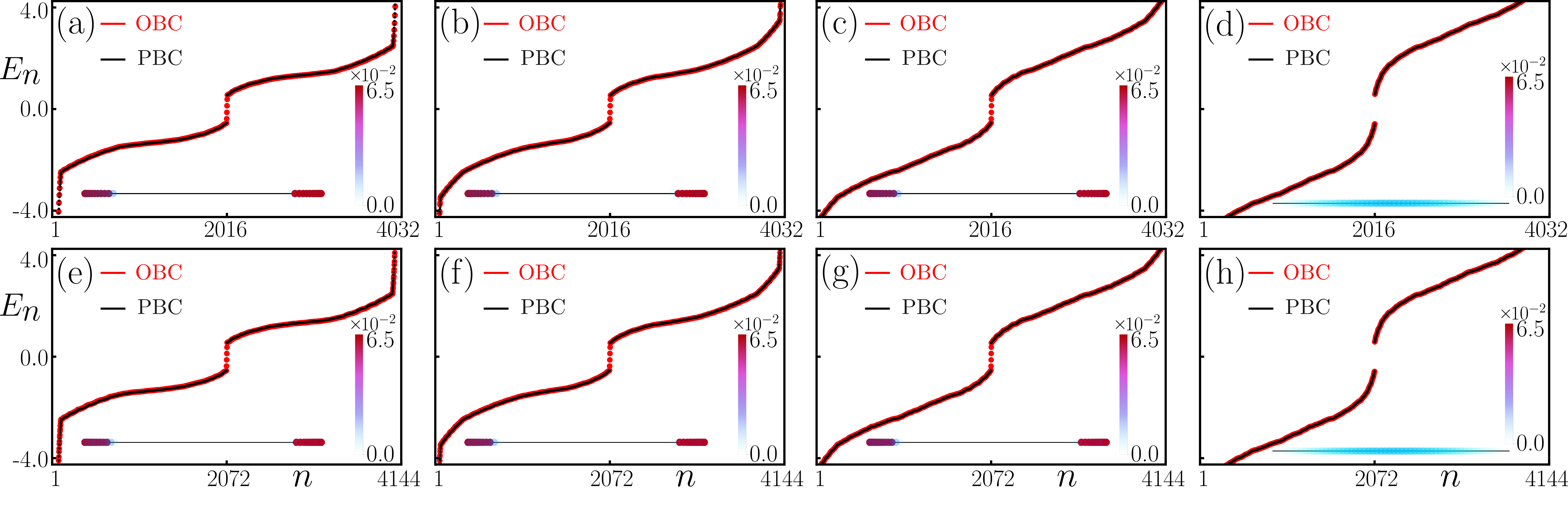}
	\caption{Energy spectrum of $H_{\rm PTB}$ [Eq.~\eqref{eq:projected-Hamiltonian}], obtained by projecting the real space Bogoliubov-de Gennes Hamiltonian $H^{\rm SL}_{\rm BdG} \equiv H_{\rm parent}$ [Eq.~\eqref{eq:Hparent}] on a $84\times 84$ parent square lattice, realized from $H_{\rm BdG}(\vec k)$ [Eq.~\eqref{eq:topo-SC-Hamtiltonian}] via Fourier transformations, onto a one-dimensional brane with an irrational slope of $m=(\sqrt{5}-1)/2$ therein (yielding a Fibonacci quasicrystal containing only $14.3\%$ sites of the parent square lattice) for (a) $m_0=0$ (a weak TSC), (b) $m_0=1$ (a strong TSC with ${\mathcal C}=-2$), (c) $m_0=2$ (a strong TSC with ${\mathcal C}=-1$), and (d) $m_0=3$ (a trivial paired state). Here, $E_n$ is the energy eigenvalue and $n$ is the eigenvalue index. Consult Fig.~\ref{fig:SC-phase-diagram} for reference. The probability density or the local density of states for two closest to zero energy states in each case is shown in the respective inset. (e)--(h) are analogous to (a)--(d), respectively, but for a projected brane with a rational slope of $m = 2/3$ within the same parent square lattice (yielding a rational approximant of the Fibonacci quasicrystal, containing only $14.7 \%$ of the parent square lattice sites). Throughout, we set $t=t_0=2 \Delta=1$ and $\mu=0$, and both types of branes are constructed with $c_\text{down} = 4$ and $c_\text{up} = 16$ [Eq.~\eqref{eq:lines}].}
	\label{fig:strong-topo-LDOS-energy}
\end{figure*}

Next, we discuss different topological phases appearing in the phase diagram of this Hamiltonian. When the chemical potential $\mu=0$, under a unitary transformation by $U = \exp(-i\frac{\pi}{4}\Gamma_{20})$, the Bloch-BdG Hamiltonian $H_{\rm BdG}(\vec k)$ becomes block-diagonal, where each two-dimensional block represents a copy of the Qi-Wu-Zhang or spinless Bernevig-Hughes-Zhang model~\cite{Das2023}. Each such copy can individually yield Chern numbers $0$ and $\pm1$, which can be obtained by implementing the Fukui-Hatsugai-Suzuki method discussed in detail in the previous section~\cite{Fukui2005} (Sec.~\ref{sec:projCTI}). When either of the blocks has a nonzero Chern number in the ground state and their sum is nonzero, the resulting superconductor is called a strong TSC, which can have Chern numbers $\pm 2$ and $\pm 1$, see the phase diagram in Fig.~\ref{fig:SC-phase-diagram}(a). It is also conceivable that such blocks have opposite (nontrivial) Chern numbers and the resulting paired state possesses net zero Chern number. Although such a paired state is devoid of any strong topological invariant (the net Chern number in this case), it features a nontrivial weak invariant, namely the Zak phase~\cite{Das2023, Zak1989}. We name this phase weak TSC, which is observed in the diamond-shaped central region (pink shaded region) in the phase diagram in Fig.~\ref{fig:SC-phase-diagram}(a). Such a phase stems from the simultaneous inversion of the BdG bands at the $\Gamma$ and $M$ points of the square lattice BZ, for which the resulting Chern numbers are equal in magnitude but opposite in sign. Consequently, despite possessing a net zero Chern number, the weak TSC is translationally active (due to the inversion of the BdG bands at the $M$ point). Finally, there exists paired state for which both the strong and weak topological invariants are zero, representing a topologically trivial paired state. Introduction of the chemical potential ($\mu$) yields scattering between these two two-dimensional blocks for $\mu=0$ and thus finite $\mu$ does not introduce any new topological paired state in the phase diagram. Finite $\mu$ only changes the parameter regime in the $(m_0,\Delta)$ plane over which various paired states can be realized, discussed next.

Cuts of the phase diagram for $H_{\rm BdG}(\vec k)$ in the $(m_0,\Delta)$ plane for different finite values of the chemical potential ($\mu$) are also calculated using the Fukui-Hatsugai-Suzuki method~\cite{Fukui2005}, discussed in detail in Sec.~\ref{sec:projCTI} and they are shown in Figs.~\ref{fig:SC-phase-diagram}(b)--\ref{fig:SC-phase-diagram}(d). Notice that for finite $\mu$ the BdG Hamiltonian $H_{\rm BdG}(\vec k)$ cannot be cast into block-diagonal form and we need to compute the total Chern number of two filled BdG valence bands at negative energies. We find that with increasing chemical potential $\mu$ the parameter regime on the $(m_0,\Delta)$ plane available for TSCs with the Chern number $\pm 2$ shrinks, as then the bands with different Chern numbers overlap (controlled by the magnitude of $\mu$) and these phases cease to exist beyond a threshold value of $\mu$ due to strong scattering between two two-dimensional blocks realized for $\mu=0$. Above such critical value of $\mu$, the phase diagram only supports strong TSCs with ${\mathcal C}=\pm 1$ and the weak TSC phase, as shown in Fig.~\ref{fig:SC-phase-diagram}(d), besides the trivial one. We also note that with increasing $\mu$, the parameter regime for the weak TSC phase shrinks and that for the strong TSCs with ${\mathcal C}=\pm 1$ increases. Therefore, for sufficiently large $\mu$, we expect the phase diagram in the $(m_0,\Delta)$ plane to only accommodate strong TSCs with ${\mathcal C}=\pm 1$ and a trivial paired state. In this work, we, however, do not consider such a large chemical doping regime. Next, we set out to harness these phases on one-dimensional PTBs at zero and moderate chemical potential, identify the corresponding bulk topological invariants (strong and weak), and responses to dislocation lattice defects.

\subsection{Strong and weak TSCs on 1D branes}

Upon Fourier transforming $H_{\rm BdG}(\vec k)$, we implement this BdG Hamiltonian from Eq.~\eqref{eq:topo-SC-Hamtiltonian} on a square lattice yielding $H^{\rm SL}_{\rm BdG}$ and subsequently project it onto one-dimensional branes to obtain the effective BdG Hamiltonian therein ($H_{\rm PTB}$) following the general principle of construction outlined in Sec.~\ref{subsec:construction}. We find that in the topological regime of $H_{\rm BdG}(\vec{k})$ (see Fig.~\ref{fig:SC-phase-diagram}), the projected BdG Hamiltonian hosts in-gap modes near zero-energy under OBCs, which are otherwise absent under PBCs. These outcomes are shown in Figs.~\ref{fig:strong-topo-LDOS-energy}(a)--\ref{fig:strong-topo-LDOS-energy}(c) for the Fibonacci quasicrystal, and their corresponding insets show the probability density or the LDOS of two near zero-energy modes confirming their strong localizations at the end points of the one-dimensional PTBs. We arrive at qualitatively similar outcome when the PTB is a rational approximant of the Fibonacci quasicrystal, as shown in Figs.~\ref{fig:strong-topo-LDOS-energy}(e)--\ref{fig:strong-topo-LDOS-energy}(g) and in their insets. By contrast, the energy spectrum is identical under PBCs and OBCs in the topologically trivial superconducting phase. As there are no edge modes on a square lattice in such a paired state, concomitantly, no endpoint zero-energy modes are found on the projected branes, as shown in Figs.~\ref{fig:strong-topo-LDOS-energy}(d) and~\ref{fig:strong-topo-LDOS-energy}(i) for the Fibonacci quasicrystal and its rational approximant, respectively. In this case, the probability density or the LDOS associated with two closest to zero-energy modes spreads throughout the bulk of the one-dimensional branes. Altogether, these observations strongly suggest that the projected BdG Hamiltonian on the effectively one-dimensional branes realizes both two-dimensional strong and weak TSCs, which we further anchor by computing the corresponding local topological invariants on the branes next.

\begin{figure}[t!]
	\centering
	\includegraphics[width=1.00\linewidth]{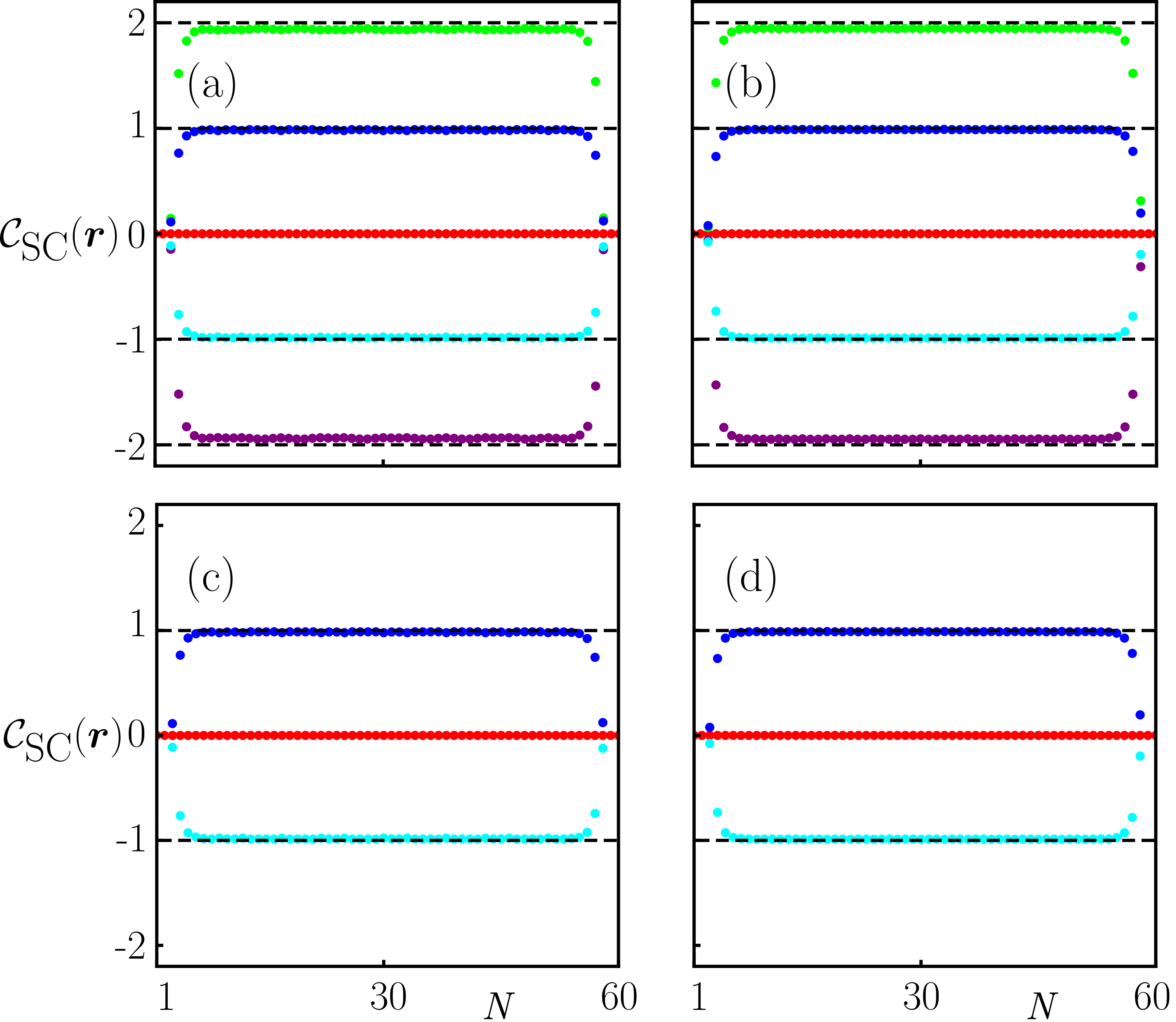}
	\caption{(a) Local Chern marker $\mathcal{C}_{\rm SC}(\vec r)$ [Eq.~\eqref{eq:local-chern-marker-TSC}] for the projected Bogoliubov-de Gennes Hamiltonian on the central line (Fig.~\ref{fig:lattice}) of the projected brane with an irrational slope of $m=(\sqrt{5}-1)/2$ within the parent square lattice (yielding a Fibonacci quasicrystal) for the pairing amplitude $\Delta=0.5$ and a chemical potential $\mu=0$. The green, blue, red, cyan, and purple curves correspond to $\mathcal{C}_{\rm SC}(\vec r)$ for $m_0 = 1,2,0,-2$, and $-1$, respectively, which are in agreement with the Chern numbers for these phases on the parent square lattice, shown in Fig.~\ref{fig:SC-phase-diagram}(a). (b) Plots of the local Chern marker with the same parameters as in (a), but on a brane with a rational slope of $m=2/3$ within the parent square lattice (yielding a rational approximant of the Fibonacci quasicrystal). (c) Same as (a), but with a finite chemical potential $\mu=1.2$, such that the $\mathcal C = \pm 2$ phases cease to exist. See Fig.~\ref{fig:SC-phase-diagram}(d) for comparison. The blue, red, and cyan curves correspond to $\mathcal{C}_{\rm SC}(\vec r)$ for $m_0 = -1, 0$, and $1$, respectively. (d) Same as (b) but for $\mu=1.2$. In all the cases, we start with a $60 \times 60$ parent square lattice and choose $c_\text{down} = 4$ and $c_\text{up}=10$, such that for the brane with irrational (rational) slope therein contains only $10.00\%$ ($10.55\%$) of its sites. Here, $N$ is the site number on the projected topological brane along the central line.}
	\label{fig:strong-topo-local-marker}
\end{figure}

\subsection{Local Chern marker for strong TSC}

Since strong TSCs possess net nonzero integer Chern numbers, we can probe it by computing the local Chern marker on the sites of the central line (see Fig.~\ref{fig:lattice}) of the projected branes~\cite{Bianco2011,Chen2023UniversalTopoMarker}, which is now given by (see Sec.~\ref{subsec:LCMTCI} for details)
\begin{equation}~\label{eq:local-chern-marker-TSC}
\begin{aligned}
    \mathcal{C}_{\rm SC}(\vec r) & = \bra{\vec r} \mathcal{L}_{\text{Chern}} \ket{\vec r} \\
                                 & \equiv \bra{\vec r} \left( -4\pi {\rm Tr}_{\{\eta,\tau\}}\left[\rm{Im}\{\hat{P} \hat{X} \hat{Q} \hat{Y} \hat{P}\}\right] \right) \ket{\vec r}, 
\end{aligned}
\end{equation}
where $\hat{P}$ is the projection operator for the two valence bands and $\hat{Q} = 1 - \hat{P}$ is the projection operator for two conduction bands for neutral BdG quasiparticles. Here, ${\rm Tr}_{\{\eta,\tau\}}$ denotes the trace over the particle-hole or Nambu ($\eta$) and the orbital ($\tau$) degrees of freedom. We find that the local Chern marker $\mathcal{C}_{\rm SC}(\vec r)$ on the sites of the projected branes near its central line (see Fig.~\ref{fig:lattice}) corroborates the quantized values of the Chern number obtained from the phase diagram of the parent square lattice-based system (see Fig.~\ref{fig:SC-phase-diagram}). In Fig.~\ref{fig:strong-topo-local-marker}, we present the local Chern marker on the sites closest to the central line of such branes with irrational and rational slopes for each value of the coordinate $x$ for various paired phases, realized on a square lattice with different Chern numbers. Indeed, we find that as the chemical potential is increased in the system the TSCs with local Chern marker $\pm 2$ disappear from the branes beyond a threshold value of $\mu$, in complete agreement with the phase diagram for the parent square lattice system, shown in Fig.~\ref{fig:SC-phase-diagram}. Once again we find that near the end points of the one-dimensional brane, the local Chern markers deviate from their quantized values found in its bulk. On the other hand, the local Chern marker assumes a trivial value for both weak TSC and trivial paired state. Next, we define and compute a local topological marker to identify such a weak TSC phase and distinguish it from the trivial paired state for which both the markers vanish. 

\begin{figure}[t!]
	\centering
    \includegraphics[width=1.00\linewidth]{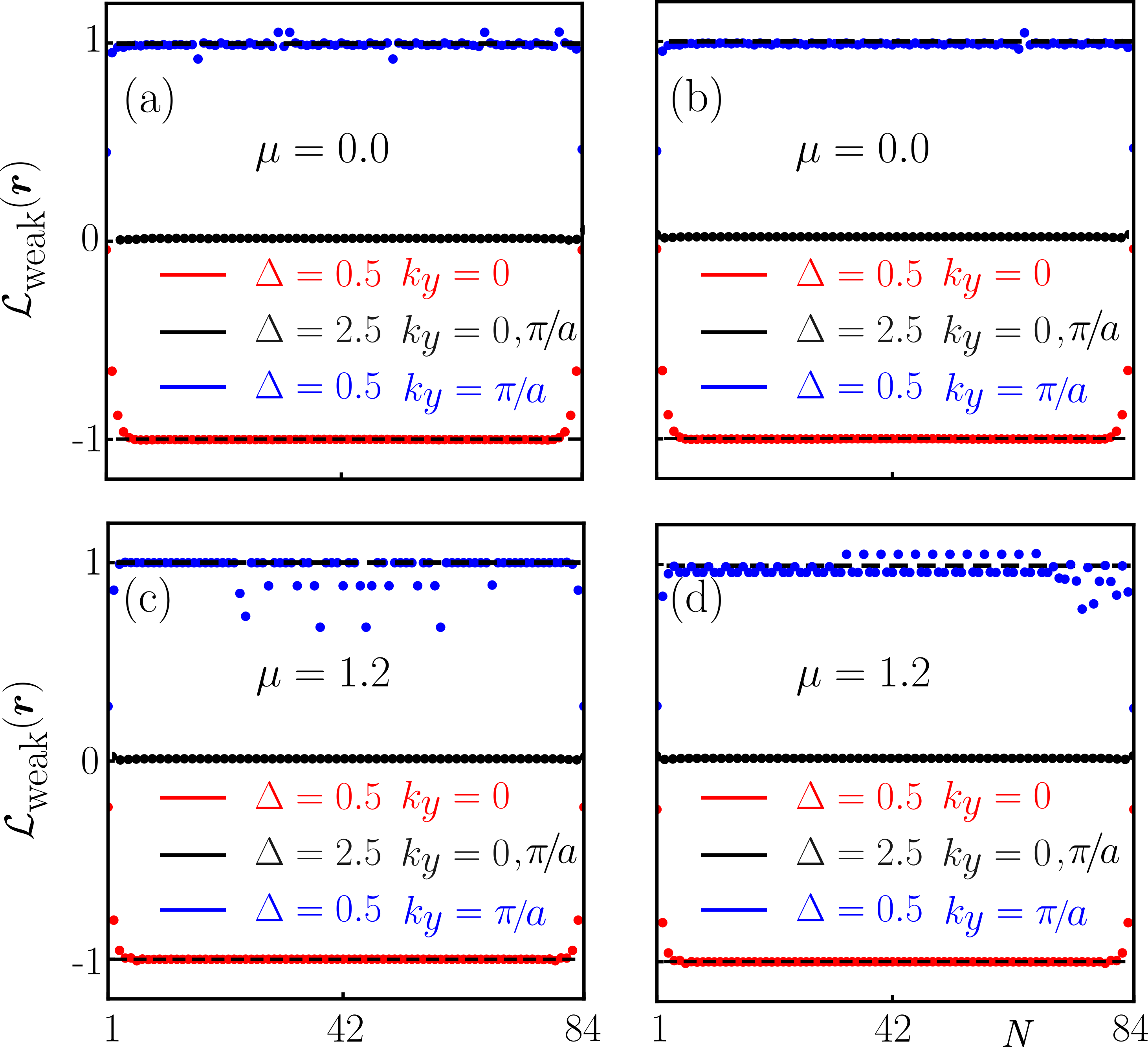}
	\caption{(a) Local weak topological marker $\mathcal{L}_{\text{weak}}(\vec r)$ [Eq.~\eqref{eq:local-weak-marker}] of the projected BdG Hamiltonian on (close to) the central line (Fig.\ref{fig:lattice}) of a one-dimensional projected brane with an irrational slope of $m=(\sqrt{5}-1)/2$ within the parent square lattice, yielding a Fibonacci quasicrystal, for $t=1$, $m_0=-0.2$, $\mu=0$, and $\Delta=0.5$ upon setting $k_y=0$ ($k_y=\pi/a$) in the Bloch-BdG Hamiltonian, showing its quantization in the weak TSC phase, except for a few sparse points when $k_y=\pi/a$. In comparison, the black curve for $\Delta = 2.5$ describes the trivial superconductor with zero local weak invariant for $k_y=0$ and $\pi/a$. (b) is same as (a), but for a brane with a rational slope of $m=2/3$ within the parent square lattice, yielding a rational approximant of the Fibonacci quasicrystal. (c) and (d) are analogous to (a) and (b), respectively, but for $\mu=1.2$. Here, $N$ is the site number on the projected topological brane along or near the central line. All the calculations are performed by starting with a $84 \times 84$ parent square lattice and setting $c_\text{down} = 6$ and $c_\text{up} = 12$ [Eq.~\eqref{eq:lines}] therein to construct the branes. Then the projected brane with a irrational (rational) slope within the parent square lattice contains $14.29\%$ ($14.68\%$) of its sites.}
	\label{fig:weak-topo-local-marker}
\end{figure}

\begin{figure*}[t!]
	\centering
    \includegraphics[width=0.96\linewidth]{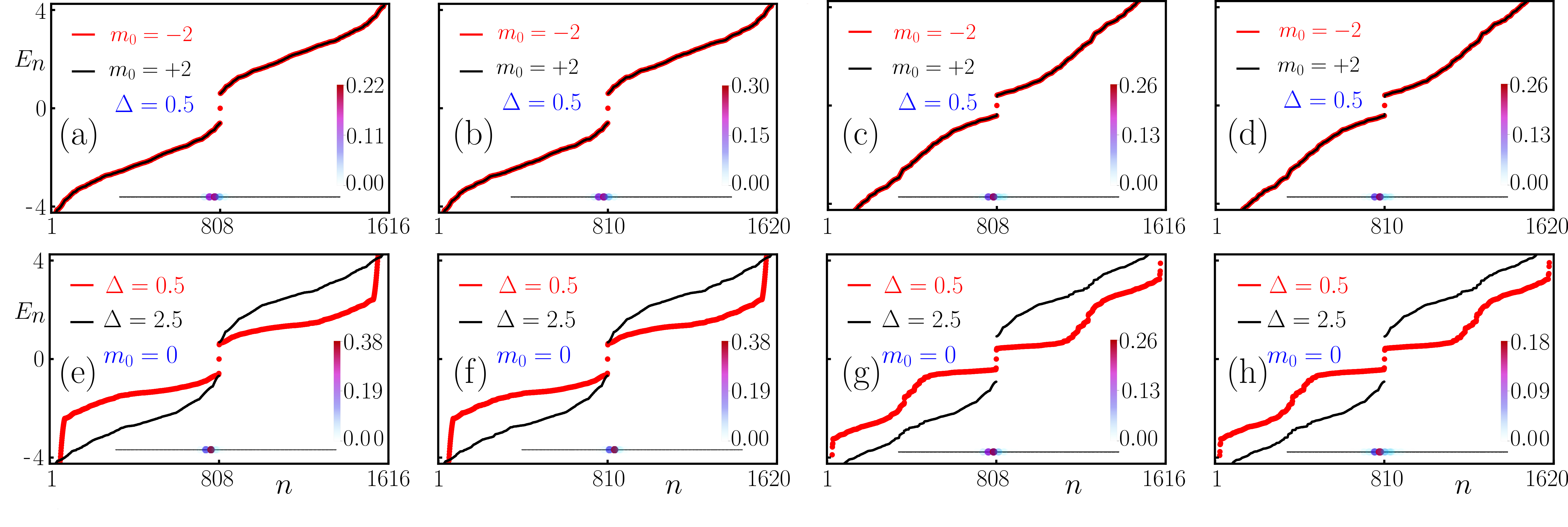}
	\caption{Responses of projected topological superconducting branes to dislocation lattice defects. (a) The energy eigenvalues ($E_n$) of the projected Bogoliubov-de Gennes Hamiltonian $H_{\rm PTB}$ of the strong topological superconductor as a function of the eigenvalue index $n$ on a projected brane with a slope of $m = (\sqrt{5}-1)/2$ within the parent square lattice (yielding a Fibonacci quasicrystal) for $m_0 =2$ (black points) and $m_0 =-2$ (red points) when $\mu = 0$. Only for $m_0 = -2$, the projected brane hosts zero energy dislocation modes, describing a translationally active paired state. (b) is analogous to (a) but for a one-dimensional brane with a rational slope of $m = 2/3$ within the parent square lattice (yielding a rational approximant of the Fibonacci quasicrystal). (c) and (d) are analogous to (a) and (b), respectively, displaying dislocation modes but for a finite chemical potential $\mu=1.2$. (e) demonstrates that the weak TSC with $m_0 = 0$ and $\Delta = 0.5$ (red points) on a Fibonacci quasicrystalline brane hosts a dislocation mode that is absent in the trivial superconductor realized for $m_0 = 0$ and $\Delta = 2.5$ (black points). (f) has same parameters as (e), but for a rational approximant brane of the Fibonacci quasicrystal. (g) and (h) are analogous to (e) and (f) but for a finite chemical potential of $\mu = 1.2$. Throughout, we construct the projected one-dimensional brane with a irrational (rational) slope within a $60\times 60$ parent square lattice with $c_{\rm up} = 2$ and $c_{\rm down}=-5$ ($c_{\rm up} = 0$ and $c_{\rm down}=-7$) such that it contains $11.20\%$ ($11.25\%$) of the original sites.}
	\label{fig:topo-SC-dislocation-LDOS-energy}
\end{figure*}

\subsection{Local topological invariant of weak TSC}

To characterize the weak topological regime for the paired state, devoid of any invariant suitable for a two-dimensional strong topological phases, such as the Chern number, Bott index, and local Chern marker, we need a different invariant, namely local weak topological marker. For that, we consider $k_y$ a good quantum number and implement the real-space hoppings in the $x$ direction. We first focus on the $k_y =0$ point, and therefore set $\sin(k_ya) \; \Gamma_{02}=0$ and $\cos(k_ya) \; \Gamma_{03}=\Gamma_{03}$. Then, following Ref.~\cite{Chen2023UniversalTopoMarker}, we construct the local weak topological marker, complementing the Zak phase in the reciprocal space~\cite{Das2023, Zak1989},  to characterize the weak TSC phase. It is given by 
\begin{equation}~\label{eq:local-weak-marker}
\begin{aligned}
	\mathcal{L}_{\rm weak}(\vec r) &= \bra{\vec r} \mathcal{L}_{{\rm weak}} \ket{\vec r} \\
    &= \bra{\vec r} {\rm Tr}_{\{\eta,\tau\}}[\hat{W} (\hat{P} \hat{X} \hat{Q} + \hat{Q} \hat{X} \hat{P})] \ket{\vec r},
\end{aligned}
\end{equation}
where $\hat{W}$ is a $4\times 4$ Dirac matrix that anticommutes with $H_{\rm BdG}(\vec k)$ with $k_y=0$ [see Eq.~\eqref{eq:topo-SC-Hamtiltonian}]. The BdG Hamiltonian in Eq.~\eqref{eq:topo-SC-Hamtiltonian} for $k_y=0$ contains the following matrices $\Gamma_{01}$, $\Gamma_{03}$, $\Gamma_{13}$, and $\Gamma_{30}$. Hence, we choose $\hat W = \Gamma_{12}$. We find that the above local weak topological marker assumes a value $-1$ in the weak TSC phase, and it is zero in the trivial paired phase on a square lattice. Recall that the Chern number (and the local Chern marker) is identically zero in both phases, see Fig.~\ref{fig:SC-phase-diagram}. The local weak topological marker retains the same value on the projected branes, which we compute in the following way. At first, we implement the BdG Hamiltonian $H_{\rm BdG}(\vec k)$ with $k_y=0$ on a $84 \times 84$ site square lattice on which the one-dimensional chains, stacked in the $y$ direction, are decoupled as $k_y$ is treated as a good quantum number. Subsequently, we project such square lattice-based Hamiltonian to a one-dimensional brane and find that the local weak invariant on the central line of the brane (see Fig.~\ref{fig:lattice}) is well-quantized to $-1$ on both Fibonacci quasicrystal and its rational approximant for zero and finite chemical potential when the parameter values are chosen such that the square lattice-bases system also hosts a weak TSC. These results are shown in Fig.~\ref{fig:weak-topo-local-marker}.

\begin{figure*}
    \centering
    \includegraphics[width=0.98\linewidth]{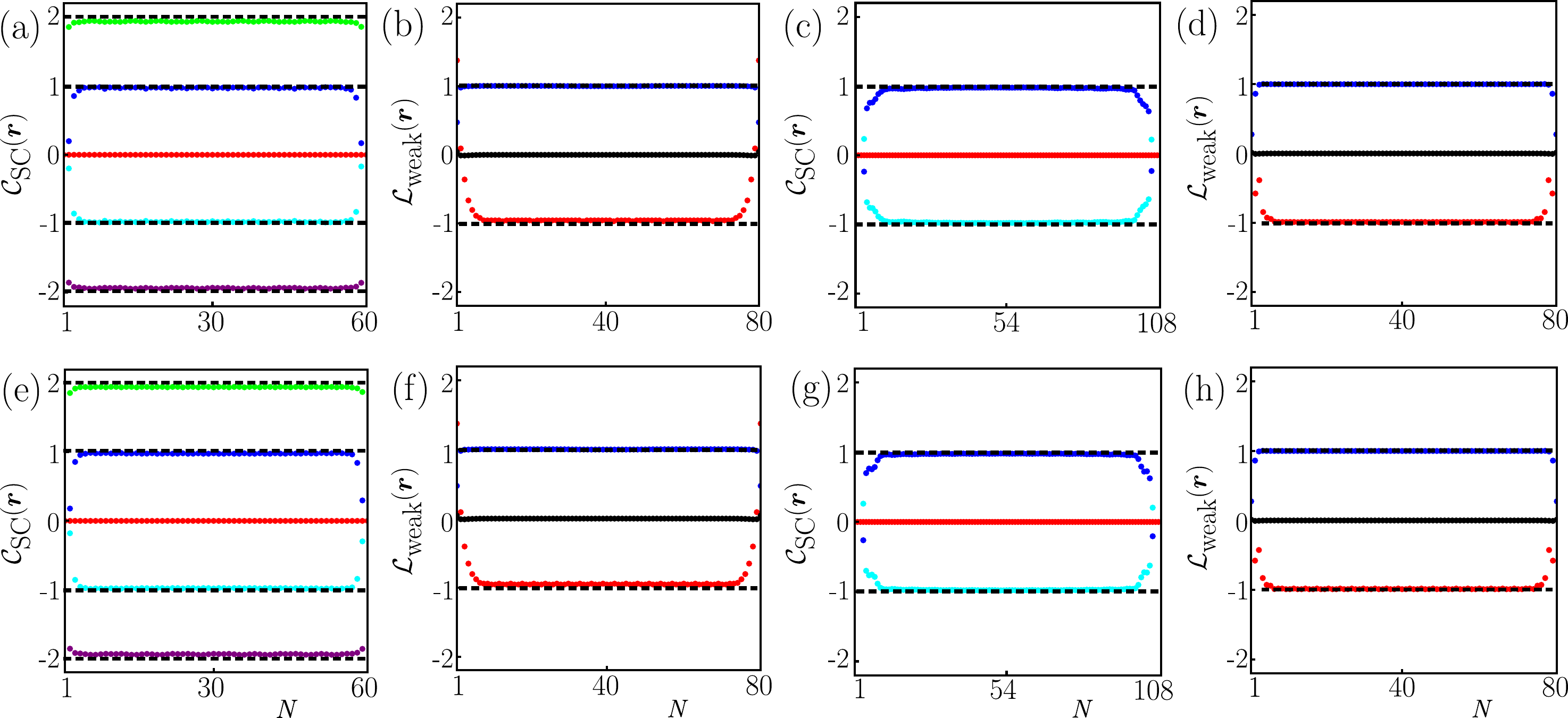}
    \caption{Local topological markers of the Bogoliubov de-Gennes (BdG) Hamiltonian constructed by projecting the Nambu-doubled normal state Hamiltonian from the parent square lattice to branes (computed along or close to its central line), and subsequently introducing the pairing and chemical potential terms therein. See Sec.~\ref{subsec:TSCcommuteTopo} for details. (a) The local Chern marker of strong topological superconductors (TSCs) on a Fibonacci quasicrystal (FQC), containing only $10.00\%$ of the parent $60 \times 60$ square lattice sites when constructed with $c_{\rm up } = 10$ and $c_{\rm down} = 4$, for $t_0=t=1$, $m_0=0$, $\mu=0$, and $\Delta=0.5$. The green, blue, red, cyan, and purple curves represent $m_0 = 1,2,0,-2$, and $-1$, respectively. (b) Local weak topological marker on a FQC, containing only $12.50\%$ of the parent $80 \times 80$ square lattice sites when constructed with $c_{\rm up } = 14$ and $c_{\rm down} = 4$, for $t_0=1, t=1, m_0=-0.2, \mu=0$. The red and blue curves are computed with $k_y=0$ and $k_y=\pi/a$, respectively, for $\Delta=0.5$ (yielding weak TSCs), and black curve is obtained for $\Delta=2.5$ (yielding a trivial pairing) and $k_y=0$ or $\pi/a$. (c) The local Chern marker of a strong TSCs on a FQC for $m_0 = -1$ (blue), $0$ (red), $1$ (cyan) at finite chemical doping of $\mu=1.2$ upon setting $t_0=t=1$ on a FQC, containing only $16.66\%$ of the $108\times 108$ parent square lattice sites. (d) Same as (b), but for chemical potential $\mu=1.2$. (e), (f), (g), and (h) are analogues to (a), (b), (c), and (d), respectively, but displaying the local topological marker of the paired state on rational approximants of the FQC, with a slope of $m=2/3$ within the parent square lattices and containing its only $10.56\%$ [in (e)], $12.91\%$ [in (f)], $16.98\%$ [in (g)], and $12.91\%$ [in (h)] of their sites.}
    \label{fig:delta-mu-commuting-order}
\end{figure*}

In order to expose the translationally active nature of such a weak TSC phase, we repeat the entire computation of the local weak topological marker by setting $k_y=\pi/a$ in $H_{\rm BdG}(\vec k)$. In this case, the local weak topological marker is quantized to $+1$ in the same weak TSC phase on majority of the sites within the branes, confirming that the inversion of the BdG bands occur simultaneously at the $\Gamma$ and $M$ points of the parent square lattice BZ, thereby yielding a net zero Chern number. But, for the trivial paired state such an invariant is always pinned to zero. These outcomes are also shown in Fig.~\ref{fig:weak-topo-local-marker}. Notice that during the computation of the weak topological invariant the $x$-directional one-dimensional (1D) chains remain decoupled in the $y$ direction, as $k_y$ is a good quantum number. Still, we need to begin with a two-dimensional lattice with no coupling in the $y$ direction on which $k_y$ is treated as a parameter, as the lattice sites constituting the projected branes reside at different $x$ coordinates, depending on the value of $y$  (because the slope of the Fibonnaci chain or its rational approximant is nonzero). We impose periodic boundary condition in the $x$ direction, while the notion of any boundary condition in the $y$ direction is moot since there is no coupling in the $y$ direction. Next, we promote the responses of both strong and weak TSCs to dislocation lattice defects.

\subsection{Dislocation modes of strong and weak TSCs}

Construction of the dislocation lattice defects on PTBs to identify translationally active phases therein has already been discussed in details in Sec.~\ref{subsec:DislocationTCI}. Here, we follow the same approach to identify translationally active strong and weak TSCs in terms of localized Majorana modes close to the zero energy near such defect cores. The momentum at which the TSCs display inversion of the BdG bands on a square lattice BZ ${\bf K}^{\rm BdG}_{\rm inv}$ can be recognized by studying $H_{\rm BdG}(\vec k)$ in the zero chemical potential limit ($\mu=0$). Then the four-dimensional BdG Hamiltonian assumes a block diagonal form, as discussed at the beginning of this section, where each such block is two-dimensional. The paired state then hosts dislocation modes as long as at least one of such blocks manifests band inversion at the $M$ point. With this criterion, we find that the strong TSC phases with the Chern number ${\mathcal C}=-1$ and $-2$ and the weak TSC are translationally active and all of them foster topologically robust dislocation modes following the ${\bf K} \cdot {\bf b}$ rule, discussed in Sec.~\ref{subsec:DislocationTCI} with ${\bf K} \equiv {\bf K}^{\rm BdG}_{\rm inv}$. By contrast, the strong TSC phases with ${\mathcal C}=1$ and $2$ are translationally inert as for them ${\bf K}^{\rm BdG}_{\rm inv}=(0,0)$ in both the blocks of the BdG Hamiltonian and they do not support any zero-energy Majorana dislocation modes. These conclusions do not change for finite chemical doping, since finite $\mu$ does not affect ${\bf K}^{\rm BdG}_{\rm inv}$.

For all the translationally active topological paired states when we project the parent square lattice-based BdG Hamiltonian with a dislocation defect onto one-dimensional branes (quasicrystalline or its rational approximant) such that the defect core resides within it, we find clear signatures of the zero-energy dislocation modes that are highly localized around its core. The corresponding results are shown in Fig.~\ref{fig:topo-SC-dislocation-LDOS-energy} at zero and finite chemical potential. Although the results are explicitly shown for the strong TSC with ${\mathcal C}=-1$ and the weak TSC with ${\mathcal C}=0$, the outcomes are qualitatively similar for the strong TSC phase with ${\mathcal C}=-2$ which we do not show in Fig.~\ref{fig:topo-SC-dislocation-LDOS-energy}  explicitly. As expected, the strong TSCs with ${\mathcal C}=1$ and $2$ do not support any zero-energy mode localized near the dislocation core on the PTBs, as the corresponding parent states on a square lattice with dislocation defects are devoid of such modes.

\subsection{Projected superconducting branes: commutative topology}~\label{subsec:TSCcommuteTopo}

Notice that the momentum-independent or local or on-site pairing yields both strong and weak topological phases in undoped or doped time-reversal symmetry breaking systems (insulator or metal) as such a paired term inherits the topology from the normal state in the following way. Namely, the normal state is described by lattice-regularized massive Dirac Hamiltonian in which the odd-parity functions $d_1(\vec{k})$ and $d_2(\vec{k})$ yields the Dirac kinetic terms in the low-energy and long wave-length limit that scale linearly with momentum. These two terms describe $p$-wave harmonics. On the other hand, $d_3(\vec{k})$ acts like a momentum-dependent Wilson-Dirac mass. The local pairing then inherits the odd-parity nature from the normal state band Hamiltonian, leading to various incarnations of the odd-parity $p$-wave pairing, which depending on various parameter values can be either topological (strong or weak) or trivial. Such a correspondence has previously been established for Bloch-BdG Hamiltonian in various dimensions and for different symmetry classes, which has subsequently been confirmed by lattice-based numerical demonstration of the bulk-boundary correspondence on regular lattices~\cite{localpairing:1, localpairing:2, localpairing:3, localpairing:4, HOTSC:11, HOTSC:15}. So far in this section, we have established that the same correspondence holds on projected superconducting branes constructed from parent crystals. Nonetheless, it is natural to raise a question regarding the fate of the topological nature of the local pairing when it is introduced after projecting the square lattice-based normal state Hamiltonian onto the one-dimensional brane.

To this end, we first implement $H_{\rm BdG}(\vec k)$ with $\Delta=\mu=0$ (Nambu-doubled normal state Hamiltonian) from Eq.~\eqref{eq:topo-SC-Hamtiltonian} on a square lattice, which we subsequently project on one-dimensional branes with irrational and rational slopes following the prescription detailed in Sec.~\ref{subsec:construction}. On such Nambu-doubled Hamiltonian for the one-dimensional projected branes, we introduce the on-site pairing and chemical potential, but only on the sites belonging to the branes. Finally, we numerically diagonalize the resulting BdG Hamiltonian to establish the bulk-boundary correspondence for the paired states in terms of the endpoint modes as well as compute the corresponding local topological markers. We find qualitatively similar results as we have discussed so far, which in turn establishes the topological robustness of the paired under the commutative nature of the projection procedure. The results are shown in Fig.~\ref{fig:delta-mu-commuting-order}.

This observation is also of importance for the realization of proposed projected TSCs in real materials. Projecting a TSC from its parent crystal and subsequently harnessing its topological properties on a lower-dimensional projected brane can be experimentally challenging to achieve. Robustness of topological properties of the local paired states under the commutative nature of the projection procedure on the other hand opens up a more realistic route to realize TSCs on projected branes in the following way. First, project the normal state onto a brane and subsequently induce the local pairing therein to realize projected TSCs. On such projected normal state branes, the desired local pairing can be externally introduced via proximity effect, for example.

\begin{figure}[t!]
    \centering
    \includegraphics[width=1.00\linewidth]{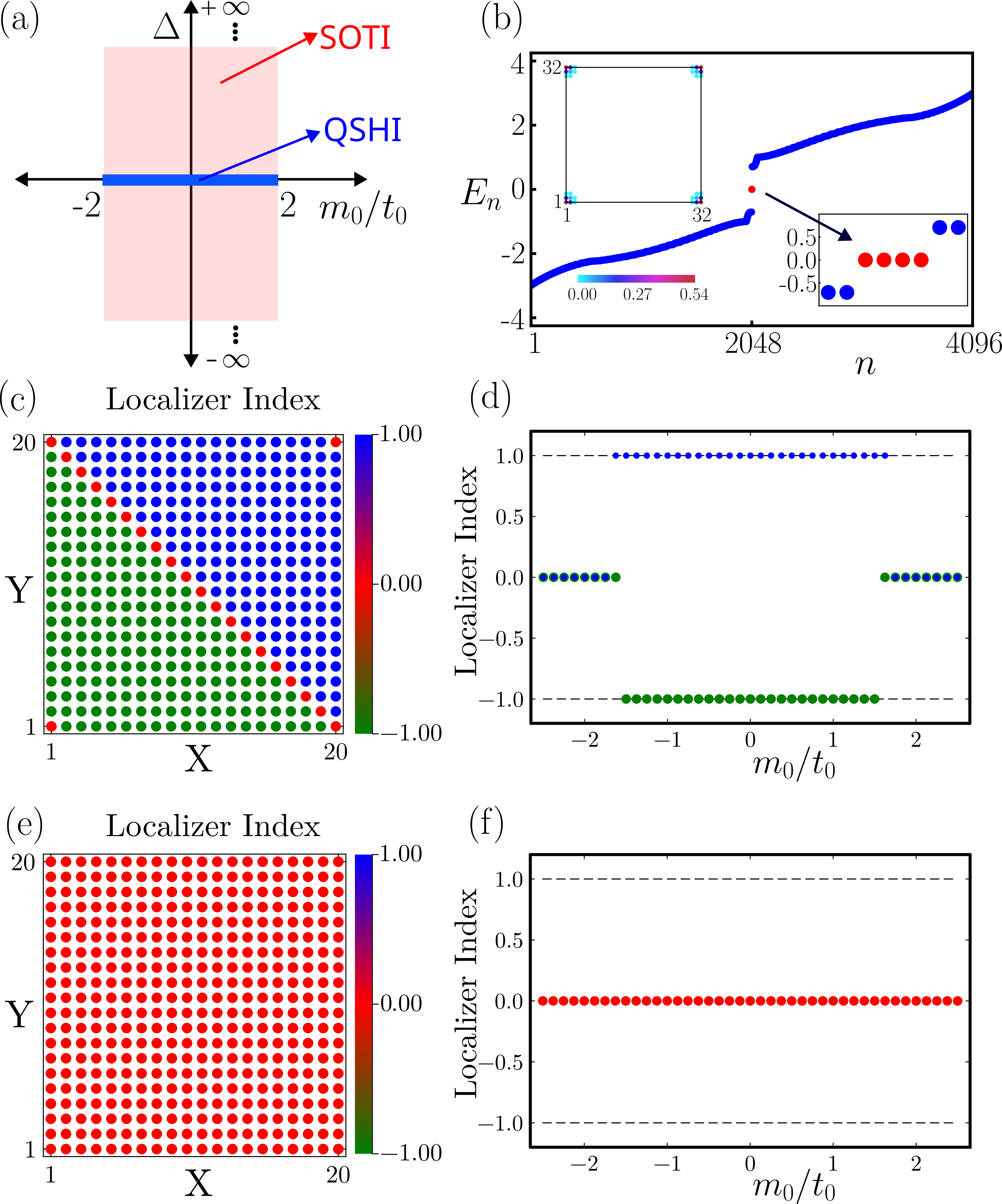}
    \caption{(a) Phase diagram of $H_{\rm SOTI}(\vec{k})$ [Eq.~\eqref{eq:HOTI-parent-hamiltonian}], hosting a second-order topological insulator (SOTI) for any $\Delta$ (pale red region) in an otherwise quantum spin Hall insulator (QSHI) phase (blue line) found for $\Delta=0$. The white region is occupied by a trivial insulator. (b) Energy eigenvalues ($E_n$) of $H_{\rm SOTI}(\vec{k})$ upon implementing it on a  $32 \times 32$ square lattice for $t = t_0 = m_0 = \Delta = 1$, where $n$ is the eigenvalue index. The insets show existence of four \emph{nearly} degenerate zero-energy corner modes (in red) and their joint local density of states. (c) The localizer index [Eq.~\eqref{eq:localizer-index}] on a $20 \times 20$ square lattice for $t = t_0 = 1, m_0=-1$ and $\Delta = 0.5$. Here the localizer index takes the quantized value of $-1$ in the bottom-left side and the quantized value of $+1$ in the top-right side. However, it takes the trivial value $0$ at the top-right and bottom-left corners besides along one of the body diagonals. (d) The localizer index on a $24\times 24$ square lattice at the points $(3,3)$ (in green) and $(21,21)$ (in blue) as a function of the parameter $m_0/t_0$ for $\Delta=1$. The localizer index takes the value $0$ in the trivial regime ($|m_0/t_0|>2$) and it takes quantized values $\pm 1$ in the topological region ($|m_0/t_0|<2$). (e) The localizer index for a QSHI with $t = t_0 = m_0 =1$ and $\Delta=0$ is identically zero everywhere in the system. (f) Variation of the localizer index over a wide range of $m_0/t_0$ for $t = t_0=1$ and $\Delta=0$ at the point $(3,3)$, encompassing both topological and trivial parameter regimes of the Bernevig-Hughes-Zhang model. 
    }~\label{fig:HOTI-parent}
\end{figure}

\section{Projected higher-order topological insulators}~\label{sec:projHOTI}

All the topological phases we have discussed so far belong to the broad family of first-order topological phases, defined in the following way. All $d$-dimensional first-order topological insulators and superconductors support topologically robust gapless modes on $d_B=(d-1)$-dimensional boundaries, which are also characterized by their codimension $d_c=d-d_B=d-(d-1)=1$. In this work, we specifically focused on the square lattice-based topological phases in $d=2$ and their boundary modes reside on the edges with $d_B=1$, thereby yielding $d_c=2-1=1$. We have demonstrated successful realizations of these two-dimensional phases (namely strong, weak, crystalline, and translationally active topological insulators and superconductors) on one-dimensional projected branes, featuring quasicrystalline or emergent crystalline structures. In this classification scheme, it has now become clear that it is possible to realize higher-order topological phases, featuring robust modes at the even low-dimensional boundaries (or on a boundary of a boundary) with $d_c=n>1$ where $n$ is an integer with corners ($d_c=d$) and hinges ($d_c=d-1$) standing as their prominent examples. In this language, an $n$th order topological phase supports gapless modes on boundaries with $d_c=n$. Such boundary modes are typically protected by composites of nonspatial and crystalline symmetries~\cite{HOT:1, HOT:2, HOT:3, HOT:4, HOT:5, HOT:6, HOT:7, HOT:8, HOT:9, HOT:10, HOT:11, HOT:12, HOT:13, HOT:14, HOT:15, HOT:16, HOT:17}, which we substantiate with a specific model in this section below. Given that projected branes are devoid of crystalline symmetries of their parent crystals, it is natural to extend the scope of our pursuit to scrutinize the fate of higher-order topological phases therein and the dependence of their boundary modes on the orientation of the branes within the parent crystal. We devote this section to answer this question by focusing on a prominent parent square lattice-based model Hamiltonian for a SOTI, supporting robust zero-energy modes that are localized at its four corners.

\begin{figure*}
    \centering
    \includegraphics[width=0.98\linewidth]{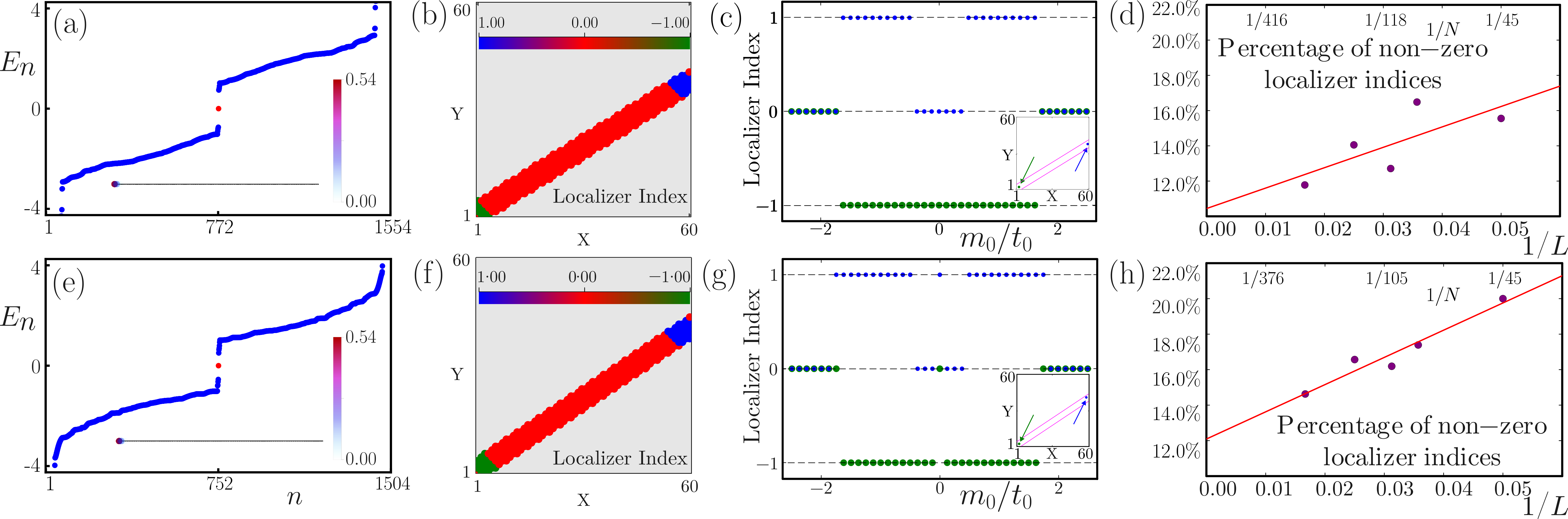}
    \caption{(a) Energy eigenvalues ($E_n$) with eigenvalue index $n$ of $H_{\rm PTB}$, obtained from 
    $H_{\rm SOTI}(\vec{k})$ for $t = t_0 = \Delta = -m_0 = 1$ [Eq.~\eqref{eq:HOTI-parent-hamiltonian}] initialized on a $60\times 60$ parent square lattice and then projected to a quasicrystalline brane (Fibonacci quasicrystal) with a slope of $m=(\sqrt{5}-1)/2$ and parameters $c_{\rm up} =4.5$ and $c_{\rm down} = -2$ [Eq.~\eqref{eq:lines}] such that only $10.72\%$ sites of the parent system are contained within the brane. Inset: local density of states of the zero-energy ($\sim 10^{-8}$) states, localized on one end of the brane that passes through one of the corners of the parent square lattice. (b) The localizer index for every site in the projected brane that is embedded within a parent square lattice (gray shaded square). While the localizer takes value 0 (red) near the middle of the brane, it still takes values $\pm 1$ (green and blue) near the ends. (c) The variation of the localization index against $m_0/t_0$ on two points with coordinates $(3,3)$ (blue) and $(59,38)$ (red) in the original square lattice that otherwise fall within the brane (as shown in the inset). (d) The dependence of percentage of sites within the brane with a nonzero localizer index on $L^{-1}$, where $L$ is the linear dimension of the parent square lattice in each direction, which saturated to $10.50\%$ (approximately) in the thermodynamic limit ($L^{-1} \to 0$). Here we have considered $L=20,28,32, 40, 60$, and ensured that the percentage of parents lattice sites within PTB remains roughly constant (between $10\%-11.5\%$). The axis on top shows the value of $1/N$, where $N$ denotes the number of sites within the brane. The parameter values in (b), (c), and (d) are same as in (a). (e)--(h) are analogues of (a)--(d), respectively, but for the rational approximant of the Fibonacci quasicrystal, constructed by taking $m=2/3$, $c_{\rm up}=4$, and $c_{\rm down} = -2$ such that it contains $10.40\%$ sites of the parent lattice. In (g), the blue and the green points denote the localizer index on the points (3,3) and (58,41), respectively, in the original square lattice that otherwise fall within the brane (as shown in the inset). In (h), the linear fit saturates to $12.50\%$ (approximately) in the thermodynamic limit. 
    }
    \label{fig:HOTI-PTB}
\end{figure*}

The Bloch Hamiltonian capturing all the quintessential properties of a two-dimensional SOTI is given by 
\allowdisplaybreaks[4]
\begin{eqnarray}~\label{eq:HOTI-parent-hamiltonian}
        H_{\rm SOTI}(\vec{k}) &=& t (\sin (k_x a) \Gamma_{31} + \sin (k_y a) \Gamma_{32}) \nonumber \\
        &+&  \big[m_0 + t_0 \{ \cos (k_xa)  + \cos (k_ya) \} \big] \Gamma_{33} \nonumber \\
        &+& \Delta \big[ \cos (k_xa) -\cos (k_ya) \big] \Gamma_{10},
\end{eqnarray}
where the four-dimensional Hermitian $\Gamma$ matrices are defined as $\Gamma_{\nu\rho} = \sigma_\nu \otimes \tau_\rho$. Two sets of Pauli matrices $\{ \sigma_\mu\}$ and $\{ \tau_\rho \}$ operate on the spin and orbital indices, respectively, with $\nu,\rho=0,1,2,3$. Notice that $H_{\rm SOTI}(\vec{k})$ contains four mutually anticommuting Hermitian Dirac matrices and their minimal representation therefore must be four-dimensional~\cite{DiracRepresentation}. The Hermitian matrix, namely $\Pi=\Gamma_{20}$ fully anticommutes with $H_{\rm SOTI}(\vec{k})$ and generates its unitary particle-hole or chiral or sublattice symmetry. With redefinitions of parameters and four mutually anticommuting Hermitian matrices, the above Bloch Hamiltonian conforms to the Benalcazar-Bernevig-Hughes model for the SOTI in two dimensions~\cite{HOT:1, HOT:11}.

The phase diagram of this model featuring higher-order topology and emergent zero-energy modes, localized at the corners of a square lattice can be appreciated in the following way. For $\Delta=0$, the above Hamiltonian mimics the Bernevig-Hughes-Zhang model, describing a quantum spin Hall insulator within the parameter regime $|m_0/t_0|<2$, which supports counter-propagating helical edge modes for opposite spin projections~\cite{classification:7}. Any finite $\Delta$ causes hybridization between such edge modes and gaps them out, except along the high-symmetry directions $k_x=\pm k_y$ along which $\cos (k_xa) -\cos (k_ya)$ vanishes and across which it changes sign. Hence, introduction of any finite $\Delta$ acts as a domain-wall mass for the counter-propagating edge modes for opposite spin projections, thereby binding zero-energy modes where it changes sign following the spirit of the Jackiw-Rebbi mechanism~\cite{HOT:10, jackiwrebbi}. Therefore, within the topological regime of the Bernevig-Hughes-Zhang model ($|m_0/t_0|<2$) any finite $\Delta$ produces a SOTI as shown in Fig.~\ref{fig:HOTI-parent}(a) and when $H_{\rm SOTI}(\vec{k})$ is implemented on a square lattice via Fourier transformations, four localized zero-energy modes appear at its four corners, as shown in Fig.~\ref{fig:HOTI-parent}(b).

Next we discuss the symmetries of $H_{\rm SOTI}(\vec{k})$ that in turn protect the corner-localized zero-energy modes of SOTI. Notice that in the $\Delta=0$ limit, the Hamiltonian, describing the Bernevig-Hughes-Zhang model, is invariant under the discrete time-reversal symmetry (${\mathcal T}$), generated by $\Gamma_{21} \; {\mathcal K}$ where ${\mathcal K}$ is the complex conjugation, inversion or parity (${\mathcal P}$) generated by $\Gamma_{33}$ under which $\vec{k} \to -\vec{k}$, and the fourfold rotational symmetry ($C_4$), generated by $\exp[i \Gamma_{03} \pi/4]$ under which $(k_x,k_y) \to (k_y,-k_x)$. The term proportional to $\Delta$ breaks all these discrete symmetries and it can thus be identified as discrete symmetry breaking Wilson-Dirac mass, which nonetheless preserves the composite $C_4 {\mathcal T}$, $C_4 {\mathcal P}$, and ${\mathcal P} {\mathcal T}$ symmetries~\cite{defect:8}. These composite symmetries in turn protect the zero-energy corner modes. Next we establish a one-to-one correspondence between the corner localized zero-energy modes and a local topological invariant for the SOTI, namely the localizer index.

\begin{figure}
    \centering
    \includegraphics[width=0.98\linewidth]{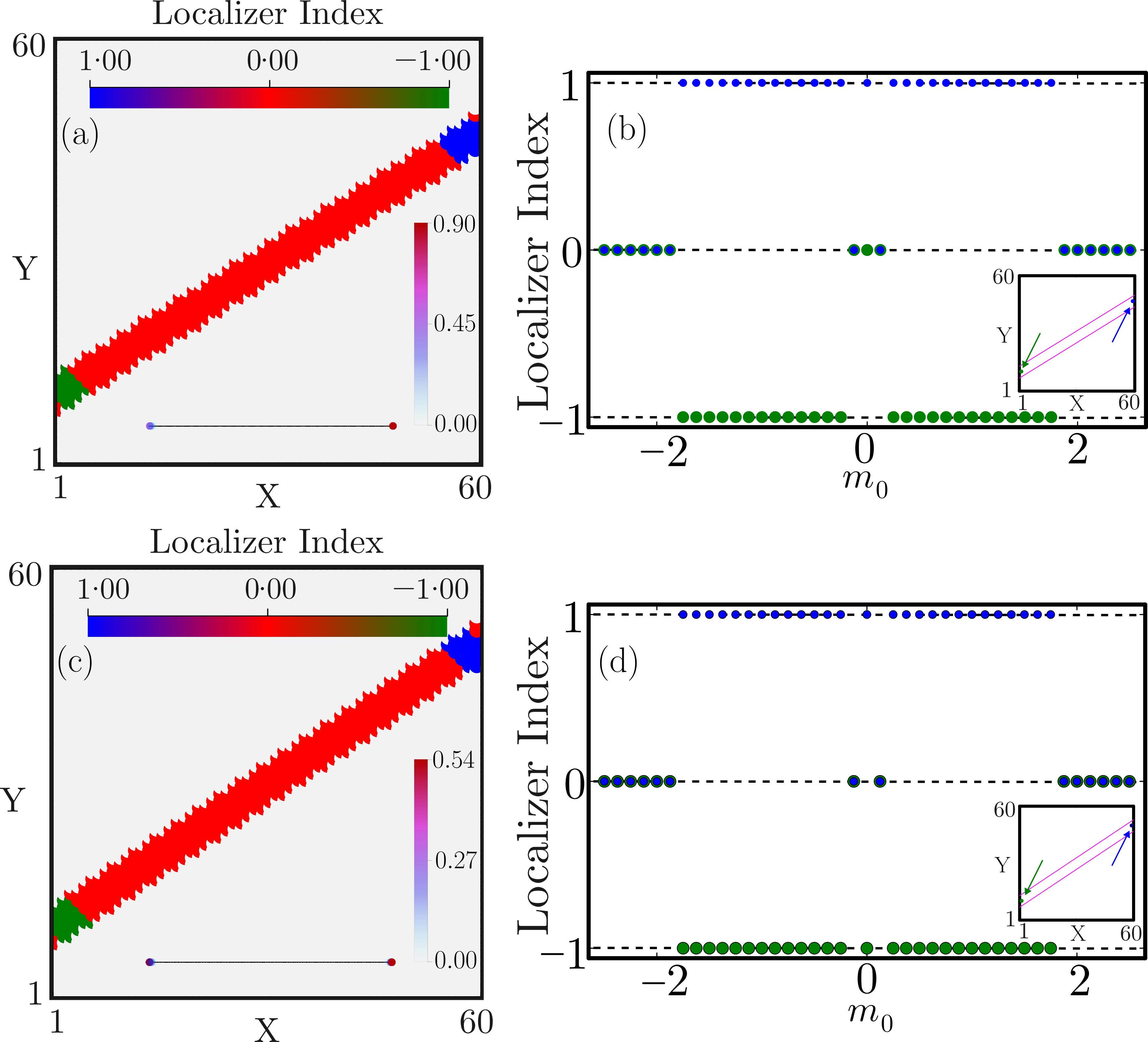}
    \caption{(a) Variation of the localizer index when the quasicrystalline brane residing at an irrational slope of $m=(\sqrt{5}-1)/2$ within the parent $60 \times 60$ square lattice contains its $10.00\%$ of sites, but its both ends are far from any corners of the parent lattice, realized by taking $c_{\rm up} = 13$ and $c_{\rm down}=7$. Other parameters are identical to Fig.~\ref{fig:HOTI-PTB}(b). (b) Variation of localizer index as a function of $m_0$ at points $(2,11)$ (green) and $(59,47)$ (blue), for a quasicrystalline brane. Other parameters are identical to Fig.~\ref{fig:HOTI-PTB}(c). (c) Same as (a), but for a rational approximant brane with slope of $m=2/3$ within the square lattice, containing its $10.55\%$ of sites. (d) Same as (b), but for the rational approximant brane, where the localizer index as a function of $m_0$ is computed at points $(2,11)$ (green) and $(59,50)$ (blue). Inset of (a) and (c) show the local density of states associated with two closest to zero-energy modes on the brane, showing their almost equal localization at two ends of the brane; compare with insets of Fig.~\ref{fig:HOTI-PTB}(a) and~\ref{fig:HOTI-PTB}(e). For the inset of (a), all the parameters are identical to those for Fig.~\ref{fig:HOTI-PTB}(a) except we take $c_{\rm up}=14$ and $c_{\rm down}=7$ in a $60 \times 60$ square lattice for which the brane contains only $11.67\%$ of parent square lattice sites. For the insets of (c), all the parameters are identical to those for Fig.~\ref{fig:HOTI-PTB}(e) and we take $c_{\rm up} = 13$ and $c_{\rm down} = 7$, such that the brane contains $10.55\%$ of parent lattice sites.   
    }
    \label{fig:HOTI-PTB-otherslope}
\end{figure}

To characterize the topological properties of a SOTI, a member of the HOTI family, we utilize the localizer index, which is applicable for both Hermitian~\cite{Cerjan2022} as well as non-Hermitian~\cite{Chadha2024} systems. To compute the localizer index for SOTI, first we define the diagonal coordinate operator $\hat{D} = (\hat{X}+\hat{Y})/{2}$ and subsequently define the localizer index at point $(x,y)$ as~\cite{Cerjan2022}
\begin{equation}~\label{eq:localizer-index}
    \nu_L(x,y) = \frac{1}{2} \mathrm{sig} \left[\left(\kappa \left(\hat{D} - \frac{x + y}{2} \right) + i H^{\rm SL}_{\rm HOTI} \right) \Pi_L \right],
\end{equation}
where $H^{\rm SL}_{\rm HOTI}$ is obtained upon implementing the Bloch Hamiltonian $H_{\rm SOTI}(\vec{k})$ from Eq.~\eqref{eq:HOTI-parent-hamiltonian} on a square lattice of linear dimension $L$ in each direction via Fourier transformations and $\Pi_L = \Gamma_{20} \otimes {\mathrm I}_L$ is the generator of the sublattice or chiral symmetry of $H^{\rm SL}_{\rm HOTI}$ that anticommutes with it where ${\mathrm I}_L$ is an $L^2$-dimensional identity matrix. The quantity $\kappa$ has an appropriate dimension such that $\kappa D$ bears the dimension of energy. We, henceforth, set $\kappa = t/a$, where $t$ is the hopping amplitude defined in Eq.~\eqref{eq:HOTI-parent-hamiltonian} and $a$ is the lattice constant. The topological signature `$\mathrm{sig}$' is defined as the difference in the number of the positive and the negative eigenvalues of the Hermitian operator residing within the square brackets in Eq.~\eqref{eq:localizer-index}. In the topological phase, the localizer index $\nu_L(x,y)$ takes values $\pm 1$ at different points of the square lattice except on one of its diagonals and at four corners, as shown in Fig.~\ref{fig:HOTI-parent}(c). Similar position dependence has also been noticed previously for an alternative real-space but global topological invariant, the quadrupole moment ($Q_{xy}$) for the SOTI phase~\cite{quadrupole:1, quadrupole:2, quadrupole:3, quadrupole:4}. These features of the localizer index hold in the entire SOTI phase, while in the trivial phase, the localizer index is identically zero for any choice of $(x,y)$, as shown in Fig.~\ref{fig:HOTI-parent}(d). With the realization of SOTI on a square lattice in terms of zero-energy corner modes and the corresponding quantized localizer index, we now proceed to search for their incarnation on one-dimensional PTBs. Finally, we note that the localizer index for the Bernevig-Hughes-Zhang model, obtained by setting $\Delta=0$ in Eq.~\eqref{eq:HOTI-parent-hamiltonian} is always identically zero as shown in Figs.~\ref{fig:HOTI-parent}(e) and~\ref{fig:HOTI-parent}(f), irrespective of the value of $m_0/t_0$. Thus, the localizer index is a real space topological invariant of solely a SOTI.

\subsection{SOTI on 1D projected branes}

To construct the Hamiltonian associated with the square lattice-based SOTI on one-dimensional PTBs, we follow the same procedure discussed so far, employed in all the previous cases. The outcomes on Fibonacci quasicrystals are shown in Figs.~\ref{fig:HOTI-PTB}(a)--\ref{fig:HOTI-PTB}(d) and those on its rational approximant are displayed in Figs.~\ref{fig:HOTI-PTB}(e)--\ref{fig:HOTI-PTB}(h). We note that when one of the end points of the PTBs passes through one of the corners of the parent square lattice we find near zero-energy modes at energies $10^{-8} \; t$ (roughly) that are highly localized at that particular endpoint, see Figs.~\ref{fig:HOTI-PTB}(a) and~\ref{fig:HOTI-PTB}(e). Associated with such endpoint mode, the localizer index assumes quantized values of $\pm 1$ on a finite fraction of sites of the PTBs, as shown in Figs.~\ref{fig:HOTI-PTB}(b) and~\ref{fig:HOTI-PTB}(f). The same conclusion holds on the PTBs for the entire parameter regime for the SOTI phase on the square lattice, see Figs.~\ref{fig:HOTI-PTB}(c) and~\ref{fig:HOTI-PTB}(g). However, we note that the localizer index equal to $-1$ shows slightly better quantization than that equal to $+1$ when compared over the entire topological parameter regime $|m_0/t_0|<2$. We believe such a difference arises from the fact that localizer index equal to $-1$ ($+1$) results from points within the brane that are close to (far from) any corner, where the hallmark corner modes are localized.

Finally, we note that as we approach the thermodynamic limit, defined as $L,N \to \infty$ with $N/L^2 \to 0$ where $L$ ($N$) is the linear dimension of the square lattice in each direction (number of sites within the PTBs), the fraction or percentage of sites on the PTBs on which the localizer index assumes quantized values saturates to a finite number, see Figs.~\ref{fig:HOTI-PTB}(d) and~\ref{fig:HOTI-PTB}(h). In turn this observation guarantees the stability of the two-dimensional SOTI phase on one-dimensional PTBs in the thermodynamic limit that in the parent square lattice is protected by the composite $C_4 {\mathcal T}$, $C_4 {\mathcal P}$, and ${\mathcal P} {\mathcal T}$ symmetries, despite the fourfold rotational $C_4$ symmetry being absent on the one-dimensional branes.

\begin{figure}
    \centering
    \includegraphics[width=0.98\linewidth]{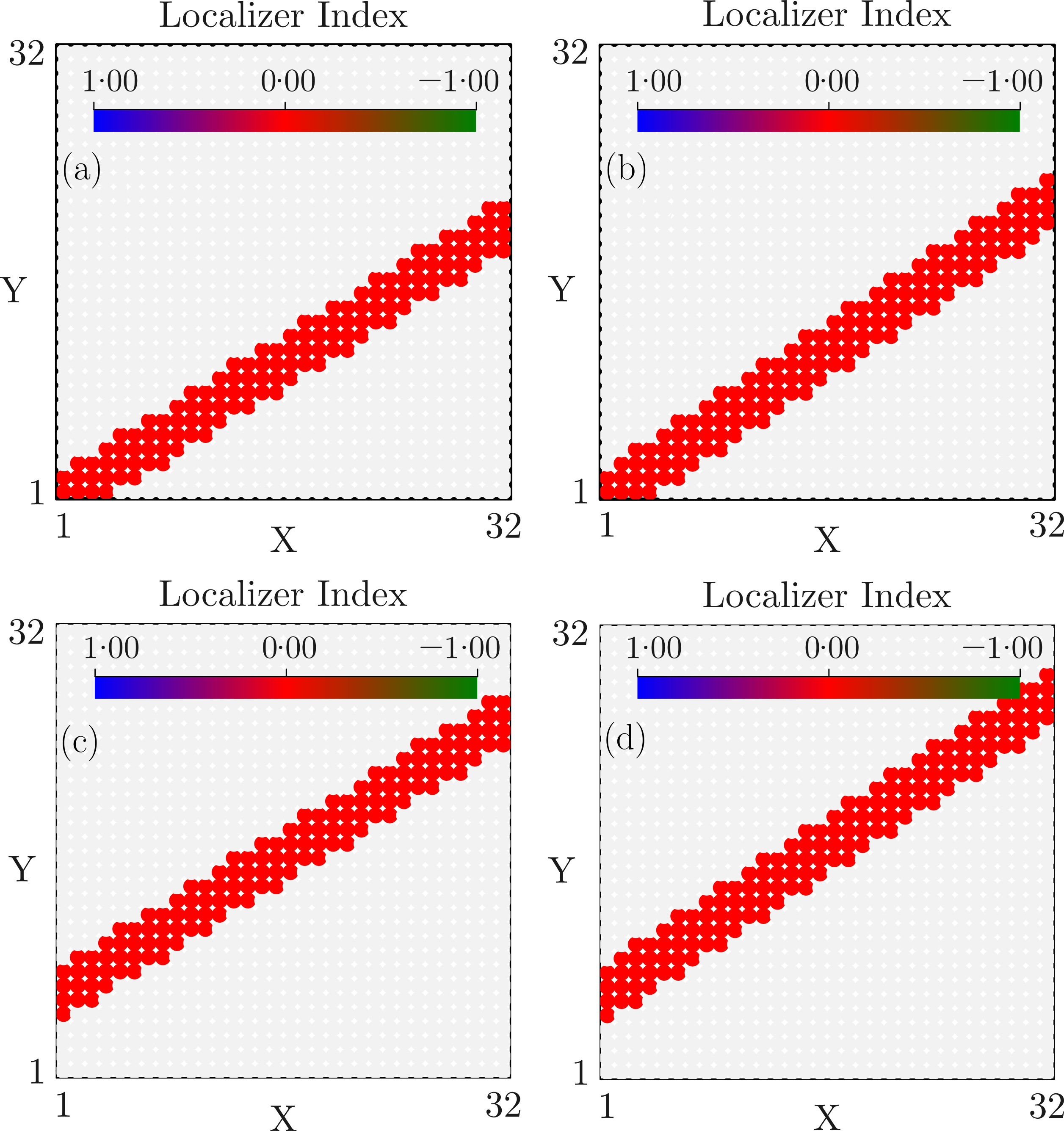}
    \caption{Localizer index for $m_0=3t$ and $t = t_0 = \Delta = 1$, demonstrating that it assumes a trivial value on the entire brane when the parameter values yield a trivial insulator on a square lattice. Throughout, we consider a $32 \times 32$ parent square lattice and project the Hamiltonian on to (a) a quasicrystalline (Fibonacci) brane with a slope $m=(\sqrt{5}-1)/2$, and {$c_{\rm up}= 2$} and $c_{\rm down} = -2$ within the square lattice, containing its $12.10\%$ of sites, and one of the ends of the brane coincides with the corner. (b) Analogous results for the rational approximant brane, with slope $m=2/3$ such that the brane contains $13.10\%$ sites of the parent square lattice. (c) Analogous plot of (a), but with $c_{\rm up} = 9$ and $c_{\rm down} = 5$ such that both the ends of the quasicrystalline brane is away from the corners of the parent lattice. Here the brane contains 12.50\% sites of the parent lattice. (d) Analogous plot of (c) with a rational slope ($m=2/3$) such that the brane contains 13.48\% sites of the parent square lattice. 
    }
    \label{fig:HOTI-PTB-trivial}
\end{figure}

We note that with two chosen slopes of the PTBs, namely $m=(\sqrt{5}-1)/2$ and $m=2/3$, yielding a Fibonacci quasicrystal and its rational approximant, it is impossible for both its ends to pass through the corners of the parent square lattice. If we choose $c_{\rm up}$ and $c_{\rm down}$ [see Eq.~\eqref{eq:lines}] such that both ends of PTBs are almost equally far from the corners of the square lattice, we find mid-gap modes at much higher energies of $10^{-2} \; t$ (roughly) for which the total probability density is almost equal at two endpoints. The behavior of the localizer index in this case is very similar to the one we discussed previously. All these outcomes are displayed in Fig.~\ref{fig:HOTI-PTB-otherslope}. Finally, we also note that the localizer index assumes trivial value everywhere on the brane, irrespective of its slope (irrational or rational) and position (passing through at least one corner or none) within the parent square lattice, when the parameter values ($|m_0/t_0|>2$) yield a trivial insulator, as shown in Fig.~\ref{fig:HOTI-PTB-trivial}. We note that recently there was an attempt to realize SOTI on Fibonacci quasicrystals, however, without any connection to any topological invariant~\cite{fibonacci:SOTI-alt}.

Finally, a comment on the quantization of the localizer index on the second-order PTB and its comparison with that on the parent square lattice is due. We note that in the entire topological regime of the parent square-lattice based SOTI phase, the PTB also features quantized localizer index but only when it is computed on a set of points that resides close to a corner of the parent square lattice. However, the localizer index computed on a set of points that resides far from one such corner can show a trivial value, but only near the singular point at $m_0/t_0=0$. These features are shown in Figs.~\ref{fig:HOTI-PTB}(c) and~\ref{fig:HOTI-PTB}(g). On the other hand, when both the ends of the PTB reside a bit far from any corner of the parent square lattice, the localizer index acquire trivial value near such a singular point irrespective of the choice of the point where localizer index is computed, as shown in Figs.~\ref{fig:HOTI-PTB-otherslope}(b) and (d). These observations are compatible with the localization and energies of the projected near-zero energy corner modes on the PTBs.

\section{Summary and discussions}~\label{sec:summary}

By considering simple yet prominent model Hamiltonian on a square lattice, describing topological insulator protected by crystalline symmetries, strong and weak topological superconductors, and second-order topological insulators, we showcase their successful incarnations on one-dimensional projected topological branes that can feature either quasicrystalline order (specifically the Fibonacci quasicrystal) or emergent periodic structure (rational approximants of the Fibonacci quasicrystal). These findings in the context of crystalline and higher-order topological insulators are of deep fundamental importance as they are protected by discrete rotational and composites of crystalline and nonspatial symmetries on the parent square lattice, respectively, which are absent in the projected branes. We also find that the local pairing yielding nontrivial topology in the paired state retains all its salient features when we introduce it after projecting the normal state Hamiltonian onto the branes. This observation is particularly important for the realization of topological paired states in lower-dimensional normal-state projected branes via the proximity effect.

While the topological crystalline insulators and superconductors (strong and weak) as a hallmark of the bulk-boundary correspondence display zero-energy modes of charged and neutral Majorana fermions at two endpoints of projected branes, respectively, the projected higher-order topological insulator strictly supports zero-energy mode at only one of its endpoints that passes through a corner of the parent square lattice. All these topological modes corroborate appropriate local strong or weak topological markers, such as the local Chern marker, local weak invariant, and the localizer index. Furthermore, when the parent first-order topological phases feature inversion of bands for charged or neutral BdG fermions at a finite momentum in the parent square lattice systems, dislocation lattice defects with their core falling within the projected branes can be utilized to identify such finite-momentum band inversion in terms of zero-energy modes localized near the center of such crystal defects. Altogether with successful realizations of strong, weak, crystalline, and higher-order topological phases of parent higher-dimensional crystals on their lower-dimensional projected branes, such lower-dimensional branes emerge as a promising material platform to harness topological phases that are realizable only in dimensions larger than three within our physical three-dimensional world.  

\begin{figure*}[t!]
    \centering
    \includegraphics[width=1.00\linewidth]{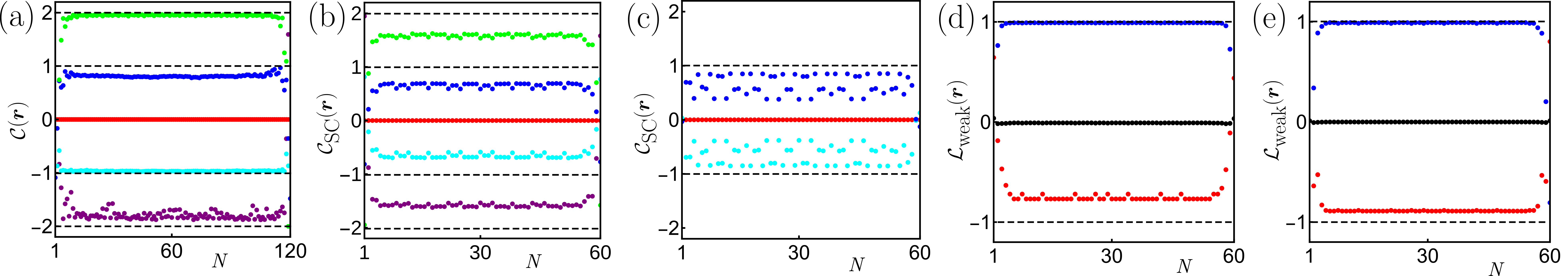}
    \caption{Local topological markers computed from $H_{11}$ only [Eq.~\eqref{eq:projected-Hamiltonian}] for the same geometry and parameter values as in (a) Fig.~\ref{fig:TCI-local-marker}(a), (b) Fig.~\ref{fig:strong-topo-local-marker}(a), (c) Fig.~\ref{fig:strong-topo-local-marker}(c), while (d) [(e)] is analogous to Fig.~\ref{fig:weak-topo-local-marker}(a) [Fig.~\ref{fig:weak-topo-local-marker}(c)] but obtained from a $60 \times 60$ parent square lattice with $c_{\rm up}=12$ and $c_{\rm down}=6$ therein. This analysis shows that by itself $H_{11}$ cannot capture all the topological properties of the branes in general and therefore systematic integration of the sites falling outside the brane is crucial to capture the topological properties of the parent crystals on their projected branes. For details see Appendix~\ref{app:comparison-H11-HPTB}.
    }~\label{app:fig:comparison-H11-HPTB}
\end{figure*}

We note that if $c_{\rm up}$ and $c_{\rm down}$ in Fig.~\ref{fig:lattice} are chosen such that two straight lines [see Eq.~\eqref{eq:lines}] encompass the entire parent square lattice, then there exists no PTB and we recover the topology of the 2D system only. On the other hand, irrespective of the values of $c_{\rm up}$ and $c_{\rm down}$, when the brane encloses only a portion of the square lattice, then it is always \emph{isomorphic} to the Fibonacci quasicrystal or its rational approximant, depending on the values of $m$ in Eq.~\eqref{eq:lines}~\cite{QCGen:2}, which then captures all the topological properties of PTBs. The key result of our paper is that even a thin, quasi one-dimensional PTB with a small value of $c_{\rm up} - c_{\rm down}$ hosts topological properties of the parent two-dimensional system, for crystalline TIs, strong and weak TSCs, and SOTI phases, as was earlier demonstrated in Ref.~\cite{PanigrahiPTB2022} only for the strong and translationally active Chern insulators.

There are multiple natural and important outgrowths of the current pursuit that would further extend the reach of projected topological branes. For example, specific momentum-independent local or on-site pairing in a higher-order normal state can nurture higher-order topological superconductors that support Majorana corner and hinge modes~\cite{HOTSC:1, HOTSC:2, HOTSC:3, HOTSC:4, HOTSC:5, HOTSC:6, HOTSC:7, HOTSC:8, HOTSC:9, HOTSC:10, HOTSC:11, HOTSC:12, HOTSC:13, HOTSC:14, HOTSC:15}. Given that current investigation has established successful realization of strong and weak topological superconductors and higher-order topological insulators on projected branes, we expect them to harbor higher-order topological superconductors from appropriate local or on-site pairings as well, explicit demonstration of which we leave for a future investigation. In addition, higher-order topological insulators support robust dislocation modes typically at finite energies that are nonetheless separated from the bulk states~\cite{defect:8}. In the future we envision capturing such finite energy defect modes and its spectral flow toward zero-energy on higher-order projected topological branes. With all the major cornerstone \emph{static} topological phases of matter being realized on projected topological branes, next we seek to extend this construction in the dynamic realm to capture Floquet topological phases~\cite{Floquet:1, Floquet:2, Floquet:3, Floquet:4, Floquet:5, Floquet:6} and real time dynamics of various topological modes~\cite{defect:15} therein, for example.

Designer materials constitute a promising quantum platform on which insulating and superconducting topological branes can in principle be engineered~\cite{designer:1, designer:2, designer:3, designer:4}. On such a platform, robust endpoint and dislocation modes can be probed via scanning tunneling microscope~\cite{STM:RMP}. Classical metamaterials, such as electrical circuits~\cite{meta:1, meta:2, meta:3, meta:4, meta:5} and phononic/acoustic~\cite{meta:6, meta:7} and photonic~\cite{meta:8} lattices, constitute yet another ground on which analogues of crystalline and higher-order topological insulators can be emulated. Topological endpoint and dislocation modes on projected circuit branes can be recognized from a diverging electrical impedance therein, while those in projected phononic and photonic branes can be detected from mechanical susceptibility and two-point pump probe or reflection spectroscopy, respectively.

It should, however, be noted that the main challenge involved in any controlled design of projected branes stems from the long-range nature of hopping amplitudes in $H_{\rm PTB}$ even though the parent system Hamiltonian ($H_{\rm parent}$) involves only short-range (such as nearest-neighbor and next-nearest-neighbor) hopping processes, which is sourced by $H^{-1}_{22}$. In Appendix~\ref{app:comparison-H11-HPTB}, we show that if we completely neglect the second term in $H_{\rm PTB}$ [Eq.~\eqref{eq:projected-Hamiltonian}] that yields infinitely long range hopping in the thermodynamic limit then the brane completely loses its topological character. As longer range hopping processes have recently been engineered on
topolectric circuits~\cite{meta:5}, at least on classical metamaterials projected branes can be engineered to recognize their topological properties. On the other hand, even though topological phases in a particular dimension are truly realized in the thermodynamic limit corresponding to an infinite lattice, numerical evidences show that finite lattices containing a few hundred lattice sites are often sufficient to capture the robust boundary modes and quantized topological invariants (within desired numerical accuracy). In order to make our theoretical proposal for projected topological brane more realistic from its design perspective in quantum materials, it will be genuinely worthwhile to test the robustness of topological modes and invariants once the range of hopping in $H_{\rm PTB}$ is systematically reduced by imposing a real-space Heaviside $\Theta$ function therein. We leave this scientific adventure as a subject for a future investigation.

\acknowledgments 

A significant portion of the computation in this work was carried out with resources provided by subMIT at MIT Physics~\cite{MITserver}. We thank the authors of the JULIA programming language~\cite{bezanson2017julia}, which was used for numerical calculations. B.R.\ was supported by NSF CAREER Grant No.\ DMR-2238679 and thanks ANRF, India, for support through the Vajra scheme VJR/2022/000022. We are thankful to Vladimir Juri\v ci\' c and Christopher A.\ Leong for critical comments on the manuscript. 

\section*{Data availability}

Numerical codes and data used in this work are available in Ref.~\cite{datacode}.

\appendix

\section{Local topological markers for bare and renormalized Hamiltonian of PTBs}~\label{app:comparison-H11-HPTB}

In our construction of the Hamiltonian for the lower-dimensional PTBs, we first consider the Hamiltonian on the parent lattice and subsequently integrate out the sites that fall outside such branes. Notice that in the resulting Hamiltonian for such PTBs, namely $H_{\rm PTB}$ in Eq.~\eqref{eq:projected-Hamiltonian}, the first term ($H_{11}$) is the bare Hamiltonian for the sites within the brane which is insensitive to the existence of the other sites in the parent lattice, while the second term ($- H_{12}H_{22}^{-1}H_{21}$) results from the above mentioned systematic site integration procedure. It is, therefore, natural to raise the following question. How important is the second term in Eq.~\eqref{eq:projected-Hamiltonian} for the branes to capture salient topological properties of the parent crystal (in this case the square lattice)? We answer this question by considering the strong and crystalline topological insulators and strong and weak TSCs. Namely, we compute the appropriate local topological markers for all these phases on the projected branes with its bare Hamiltonian $H^{\rm bare}_{\rm PTB}=H_{11}$ and compare the outcomes from the renormalized Hamiltonian $H^{\rm ren}_{\rm PTB}=H_{\rm PTB}$ [Eq.~\eqref{eq:projected-Hamiltonian}], shown in Figs.~\ref{fig:TCI-local-marker},~\ref{fig:strong-topo-local-marker}, and~\ref{fig:weak-topo-local-marker}. The results are displayed in Fig.~\ref{app:fig:comparison-H11-HPTB}, showing that for all these cases the corresponding local topological markers typically deviate considerably from their expected quantized values when computed from $H^{\rm bare}_{\rm PTB}$ and quantized local topological markers in general can only be found from $H^{\rm ren}_{\rm PTB}$. This observation in turn establishes the paramount importance of constructing the effective Hamiltonian for projected branes from its parent crystal by systematically integrating out the sites falling outside it, which is equivalent to the Schur complement of the sites residing within the brane. Results in Fig.~\ref{app:fig:comparison-H11-HPTB} are shown for the Fibonacci quasicrystals, while they are qualitatively similar on their rational approximant and thus not shown explicitly.


\end{document}